\documentclass[aps,prd,preprintnumbers,showpacs]{revtex4}
\usepackage{graphicx,epsfig}
\usepackage{subfigure}
\newcommand {\bc}{\begin {center}}
\newcommand {\ec}{\end {center}}
\newcommand {\be}{\begin {equation}}
\newcommand {\ee}{\end {equation}}
\newcommand {\beq}{\begin {eqnarray}}
\newcommand {\eeq}{\end {eqnarray}}
\newcommand {\unl}{\underline}
\newcommand {\ovl}{\overline}

\def\intl {\int\limits}
\def\lbar {\lambda\hskip-5pt\raise3pt\hbox {--}}
\def\lbr {\lambda\raise2pt\hbox {\hskip-4pt{\scriptsize --}}_\C}
\def\oa {\ovl {a}}
\def\disp {\displaystyle}
\def\prt {\partial}

\renewcommand{\d}{{\rm d}}

\def\ff {{\rm f}}
\def\ii {{\rm i}}
\def\ph {{\rm ph}}
\def\ve {\textit{\textbf{e}}}
\def\vk {\textit{\textbf{k}}}
\def\vp {\textit{\textbf{p}}}
\def\vr {\textit{\textbf{r}}}
\def\vv {\textit{\textbf{v}}}
\def\vpz {\textit{\textbf{p}}_{YZ}}

\def\vN {\textit{\textbf{N}}}

\def\ue {\unl {e}}
\def\uk {\unl {k}}

\def\ur {\unl {r}}
\def\uA {\unl {A}}
\def\cA {{\mathcal A}}
\def\cB {{\mathcal B}}
\def\cF {{\mathcal F}}

\def\cS {{\mathcal S}}
\def\cN {{\mathcal N}}

\def\cR {{\mathcal R}}
\def\cV {{\mathcal V}}

\def\ugam {\unl {\gamma}}
\def\unb {\unl {\nabla}}
\def\Dr#1#2{{\frac {\prt #1}{\prt #2}}}
\def\DR#1#2{{\frac {\d #1}{\d #2}}}

\begin{document}

\title{Relativistic kinetic equation for Compton scattering of polarized radiation in strong magnetic field}

\author{Alexander A. Mushtukov $^{1,2,3}$}
\email{al.mushtukov@gmail.com}
\author{Dmitrij I. Nagirner $^1$}
\email{dinagirner@gmail.com}
\author{Juri Poutanen $^3$}
\email{juri.poutanen@oulu.fi}

\affiliation{$^1$Sobolev Astronomical Institute, Saint Petersburg State University,
  Saint-Petersburg 198504, Russia \\
$^2$Pulkovo Observatory of Russian Academy of Sciences,
  Saint-Petersburg 196140, Russia \\
$^3$Astronomy Division, Department of Physics, PO Box 3000, 
FIN-90014 University of Oulu, Finland
}

\date{\today}

\begin{abstract}
We derive the relativistic kinetic equation for Compton scattering of polarized radiation in strong magnetic field using the Bogolyubov method.
The induced scattering and the Pauli exclusion principle are taken into account.  The electron polarization   is also considered
in the general form of the kinetic equation. 
The special forms of the equation for the cases of the non-polarized electrons, the rarefied electron gas and 
the two polarization mode description of radiation are found. 
The derived equations are valid for any photon and electron energies and the magnetic field strength below about $10^{16}$ G.
These equations provide the basis for formulation of the equation for polarized radiation transport   
in atmospheres and magnetospheres of strongly magnetized neutron stars.
\end{abstract}

\pacs{52.25.Dg, 52.25.Os, 95.30.Gv, 95.30.Jx, 97.60.Jd}

\maketitle

\section{Introduction}

Observations of the soft gamma-ray repeaters and anomalous X-ray pulsars showed that these objects can be associated with the strongly magnetized 
neutron stars (NSs) with the magnetic field exceeding the Schwinger critical value of $B_{\rm cr} = m_{\rm e}^2 c^3/e\hbar  = 4.412\times 10^{13}$ G \cite{Mere02,WT06}. 
This has revived the interest in theoretical studies of the interaction processes  
between radiation and  matter in such fields \cite{HL06}.  

Compton scattering is an important process shaping the radiation spectra of the NS atmospheres.  Its properties 
in the magnetic field differs substantially from the case when the magnetic field is absent. 
Even the classical non-relativistic limit of the scattering cross section has a resonance at the energy related to the Lorentz frequency 
and is strongly dependent on the photon energy, polarization and the B-field strength \cite{Can1971, Ven1979}. 
While the classical description has been useful for understanding the approximate effects  of energy, angle and polarization dependence of the cross section in the magnetic field, it does not include the possibility of the electron excitation to a higher Landau state
corresponding to the resonances at higher harmonics, which required fully relativistic treatment.
In the relativistic regime the recoil of the electron is important and the natural line width of the cyclotron resonances depends on the spin of the electron. The relativistic scattering cross section
for the simplest case of ground-to-ground state scattering in the magnetic field  was derived in \cite{Her1979}. 
These results were extended to a more general case of scattering 
to arbitrary Landau states in \cite{DH1986, BAM1986} and discussed further in \cite{Mes1992}. 
The derived expressions have been applied to modeling the cyclotron line formation in accreting neutron star atmospheres, 
but only for the case of one-dimensional thermal electron distribution because of the complexity of the expressions 
\cite{AlMesz1989,AlMesz1991,HarDaugh1991,ArHar1996,ArHar1999}. 
When the  incident photons propagate along the magnetic field, the resonance appears only at the fundamental frequency 
and scattering to the higher Landau levels can effectively be neglected. This allows to simplify 
the expressions for the relativistic cross sections and to approximate them by  analytical formulae \cite{GHBCM2000}.

The transport of photons through the atmosphere involves multiple scattering, which have to be 
considered either by the Monte Carlo methods or using the kinetic equations. 
The former approach was used for a qualitative study  of the line formation process in Her X-1 \cite{Yah1979, Yah1980},   
but it becomes impractical for a large optical depth and when the  induced scattering has to be accounted for, and 
therefore has a limited field of applications.
In the cold plasma approximation, assuming  the coherent scattering, the radiative transfer equation 
can be formulated as a set of  coupled equations for two normal polarization modes \cite{GP1974}. 
The influence of the electron temperature on the radiation transport can be accounted by the Fokker-Planck 
approximation, for example, by modifying the Kompaneets equation \cite{K1956} 
to allow for the resonances in the scattering cross section \cite{BHP1979}. 
Such a treatment, however, does not account for the effects of the photon angular distribution and  polarization. 
Photon polarization, however, influences the photon redistribution over the energy \cite{Nag1981,PSM1989}.

In a sufficiently strong magnetic field, owing  to the large Faraday depolarization, the radiation can be described in terms of two polarization modes. 
Under certain conditions (depending on the field strength, photon energy and propagation direction), however, 
the vacuum resonance is accompanied by the phenomenon of mode collapse and the breakdown of Faraday depolarization \cite{Zhel1983,LH2003, PSh1979}. In this case the two-mode description fails and instead the kinetic equations have to written in terms of the Stokes parameters or the coherency matrix. 
In the case when the induced scattering needs to be accounted for, the situation complicates further as 
there is no intuitive way to get such an equation.

The aim of this paper is to derive from first principles a general kinetic equation for Compton scattering in any magnetic field 
accounting simultaneously for photon polarization in terms of the Stokes parameters, for the induced scattering and the Pauli exclusion 
principle for electrons. 
We use methods of quantum statistics and follow an approach similar to that used for derivation of the kinetic equation 
for Compton scattering without magnetic field  \cite{NP2001}.
The resulting equations are valid for any photon and electron energies, 
and for the magnetic field strength below about $10^{16}$ G.
In the most general case, the electron polarization is also taken into account. 
We also consider several special cases and derive the kinetic equations when 
the electron gas  is non-polarized  and rarefied as well as when the radiation can be presented via two polarization modes. 
The derived equations provide the basis  for construction of the models of  radiation transport   
in atmospheres and magnetospheres of strongly magnetized neutron stars.


\section{Description of the electron and photon gases}

We use the system of units where $\hbar=c=m_{\rm e}=1$. We 
assume that the magnetic field is locally homogeneous, which is  justified, because  
the scales of changes of the B-field are orders of magnitude larger than the microscopic 
magnetic scale for conditions in atmospheres of NS and even the geometrical depth of the atmosphere. 
The magnetic field is described by the dimensionless parameter $b=B/B_{\rm cr}$. 
We choose the reference frame in any space-point so that the $z$-axis coincides with the magnetic field direction.
The following  assumptions about the time scales are used: 
\begin {enumerate}
\item The typical time scale on which the distribution function changes (for electrons and photons) is much larger than the typical time scale between the interactions.
\item The plasma is sufficiently rarified, so that we can use a generalization of 
 the Bogolyubov method for the case of quantum statistics to derive the kinetic equation.  
\item The typical time scale of a single interaction is much smaller than the typical time scale between the interactions. 
\end {enumerate}

\subsection{Descriptions of single particles}

The electron states are described by the wave-functions $\Psi_{n\sigma}(\ur,Y,Z)$. Its arguments are the space-time coordinates,  
the momentum projection $Z$ on the direction of the magnetic field, 
$Y$ describing location of the center of electron gyro-orbit (its $y$-coordinate),
the Landau level $n$, and the spin projection $\sigma$ on the magnetic field direction ($\sigma=\pm 1$ in $\hbar/2$ units). 
Dimensionless energy of an electron in this case, 
\be \label{eq:RnZ}
R_n(Z)=\sqrt {1+Z^2+2bn},
\ee
is independent of $Y$. 
The energy levels are degenerate with the spin projection on the magnetic field direction, 
except for the ground Landau level with $n=0$, where $\sigma$ can have only one value $-1$.
The full electron wave function is presented through the partial solutions of the Dirac equation for the electron in the magnetic field:
\be \label{eq:psiurfull}
\psi(\ur)=\sum_{n,\sigma}\int\frac{\d Y\d Z}{R_n(Z)}
\Psi_{n\sigma}(\ur,Y,Z)b_{n\sigma}(Y,Z),
\ee
where $b_{n\sigma}$ are coefficients.

The photon state is described by four parameters: the wavenumber $k$, the two angles $\theta$ and $\varphi$, which define the direction of the photon momentum, and the polarization state $s=1,2$. The 3-dimensional photon momentum can be represented as
\be \label{eq:vkphoton}
\vk=(k_x,k_y,k_z)=k(\sin\theta\cos\varphi,\sin\theta\sin\varphi,
\cos\theta).
\ee
The corresponding photon 4-momentum is $\uk=\{k,\vk\},\,k=|\vk|$.
Photon polarization is described by the polarization basis. It consists of two unit vectors, which are orthogonal to the photon momentum $\vk$:
\be \label{eq:ve1ve2orts}
\ve_1=(\sin\varphi,-\cos\varphi,0),\quad
\ve_2=(\cos\theta\cos\varphi,\cos\theta\sin\varphi,-\sin\theta).
\ee
The 4-vector potential can be defined as: 
\be \label{eq:uAsur}
\uA_s(\ur)=\ue_se^{-i\uk\cdot\ur},\quad\ue_s=\{
0,\ve_s\},\quad s=1,2.
\ee
We note that the photons are described in the same manner as in the case when the magnetic field is absent. 
We assume that the dispersion relation for the photons in magnetized vacuum does not differ 
from the dispersion relation in the case when the magnetic field is absent. 
This approximation constrains the strength of the field and energies of photons. 
For estimations one needs to know vacuum dielectric tensor and the inverse permeability tensor
for the case of magnetized vacuum \cite{Adler1971,PLShH2004}.
It is known that the indices of refraction differ from unity by more than $10\%$ 
 only for the fields with strength $b>300$ \cite{ShHL1999}.  This restricts  
 application of the developed formalism to $B\lesssim 10^{16}$ G.

\subsection{Description of the particle ensembles}

\subsubsection{Wave functions}

We describe particle ensembles by density matrix using the rules of quantum statistics. Let us define the wave functions for the case of limited number of particles. These functions will be used for construction of the density matrix.
The wave function for a limited number of particles with defined characteristics of each of them, can be found from the vacuum wave function by applying the operators of creation and annihilation. 
Let $\bar{a}_{(s)}(\vk)$ and $a_{(s)}(\vk)$ be the creation and annihilation operators of a photon in the state with polarization $s$ and 3-momentum $\vk$. According to the methods of second quantization \cite{BS1959}, these operators satisfy the relation 
\be \label{eq:aoacommut}
a_{(s)}(\vk)\oa_{(s')}(\vk')-\oa_{(s')}(\vk')a_{(s)}(\vk)=k
\delta(\vk-\vk')\delta_{s}^{s'}.
\ee

Let $b^\dag_{n\sigma}(Y,Z)$ and $b_{n\sigma}(Y,Z)$ be creation and annihilation operators of an electron on the Landau level $n$ in polarization state $\sigma$ with momentum projections $Y$ and $Z$. These operators satisfy the following relation
\be \label{eq:bdbcommut}
b_{n\sigma}(Y,Z)b^\dag_{n'\sigma'}(Y',Z')+b^\dag_{n'\sigma'}(Y',Z')
b_{n\sigma}(Y,Z)=R_n(Z)\delta(Y-Y')\delta(Z-Z')\delta_{n}^{n'}
\delta_{\sigma}^{\sigma'}.
\ee
The system of $N$ photons with fixed parameters $\{s_1,\vk_1;...;s_N,\vk_N\}$ is described by the wave function
\be \label{eq:PsisvkN}
\Psi_{s_1...s_N}(\vk_1,...,\vk_N)=\frac{1}{\sqrt {N!}}\bar {a}_{(s_1)}
(\vk_1)...\bar {a}_{(s_N)}(\vk_N)\Psi_0,
\ee
where $\Psi_0$ is the vacuum wave function of the photon gas.
Analogously, the system of $N$ electrons with fixed parameters $\{n_1,\sigma_1,Y_1,Z_1;...;n_N,\sigma_N,Y_N,Z_N\}$ is described by the wave function 
\be \label{eq:PhisigYZ}
\Phi^{\sigma_1...\sigma_N}_{n_1...n_N}(Y_1,Z_1,...,Y_N,Z_N)=
\frac{1}{\sqrt {N!}}b^{\dag}_{n_1\sigma_1}(Y_1,Z_1)...b^{\dag}_{n_N
\sigma_N}(Y_N,Z_N)\Phi_0.
\ee
where $\Phi_0$ is the vacuum wave function of the electron-positron gas.
The wave function for arbitrary state of particles can be presented as a sum of the wave functions with fixed particle parameters. For example, the state of $N$ photons is described by the function
\be \label{eq:PsiNdef}
\Psi_N=\int\prod_{j=1}^N\frac{\d\vk_j}{k_j}c_{s_1...s_N}(\vk_1,...,\vk_N)
\Psi_{s_1...s_N}(\vk_1,...,\vk_N),
\ee
where $c_{s_1...s_N}(\vk_1,...,\vk_N)$ are the weight coefficients. 
The wave function for an arbitrary state of $N$ electrons can be represented as 
\be \label{eq:PhiNdef}
\Phi_N=\int\prod_{j=1}^N\frac{\d Y_j\d Z_j}{R_{n_j}(Z_j)}
c_{n_1...n_N}^{\sigma_1...\sigma_N}(Y_1,Z_1,...,Y_N,Z_N)
\Phi^{\sigma_1...\sigma_N}_{n_1...n_N}(Y_1,Z_1,...,Y_N,Z_N).
\ee
The wave function for state with $N$ photons and $N_{+}$ electrons can be written as 
\beq \label{eq:PsiNNpdef}
\strut\disp \Psi_{N,N_{+}} & = &  \int\prod_{i=1}^N\frac{\d\vk_i}{k_i}
\prod_{j=1}^{N_{+}}\frac{\d Y_j\d Z_j}{R_{n_j}(Z_j)}c_{n_1...n_{N_{+}},s_1...
s_N}^{\sigma_1...\sigma_{N_{+}}}(Y_1,Z_1,...,Y_{N_{+}},Z_{N_{+}},\vk_1,...,
\vk_N)\nonumber \\
\strut\disp & \times& \Psi_{s_1...s_N}(\vk_1,...,\vk_N)\Phi^{\sigma_1...
\sigma_{N_{+}}}_{n_1...n_{N_{+}}}(Y_1,Z_1,...,Y_{N_{+}},Z_{N_{+}}). 
\eeq

\subsubsection{Density matrix}

The density matrix is defined as an averaged dyad product of the state vector with its conjugate. 
For the system consisting of $N$ photons and $N_{+}$ electrons it can be written in the form
\beq \label{eq:rhoNNpdef}
& \strut\disp \rho_{N,N_{+}}=\frac{1}{N!N_{+}!}\int\prod_{i=1}^N
\frac{\d\vk_i}{k_i}\frac{\d\vk'_i}{k'_i}\int\prod_{j=1}^{N_{+}}
\frac{\d Y_j\d Z_j}{R_{n_j}(Z_j)}\frac{\d Y'_j\d Z'_j}{R_{n'_j}(Z'_j)}
 & \nonumber \\
& \strut\disp \times \left\langle c{^*}{_{n'_1...n'_{N_{+}},s'_1...s'_N}
^{\sigma'_1...\sigma'_{N_{+}}}}(Y'_1,Z'_1,...,Y'_{N_{+}},Z'_{N_{+}},
\vk'_1,...,\vk'_N)c_{n_1...n_{N_{+}},s_1...s_N}^{\sigma_1...\sigma_{N_{+}}}
(Y_1,Z_1,...,Y_{N_{+}}Z_{N_{+}},\vk_1,...,\vk_N)\right\rangle
& \nonumber \\
& \strut\disp 
\times\Psi_{s_1...s_N}(\vk_1,...,\vk_N)\ovl {\Psi}_{s'_1...s'_N}
(\vk'_1,...,\vk'_N)\Phi^{\sigma_1...\sigma_{N_{+}}}_{n_1...n_{N_{+}}}(Y_1,Z_1,
...,Y_{N_{+}},Z_{N_{+}})\ovl {\Phi}^{\sigma'_1...\sigma'_{N_{+}}}_{n'_1...n'_
{N_{+}}}(Y'_1,Z'_1,...,Y'_{N_{+}},Z'_{N_{+}}). &
\eeq
The expressions in the triangle brackets are the elements of the density matrix kernel.

\subsubsection{Algebra of the density matrix kernels}

From now on we will operate only with density matrix kernels. All the equations and the final results are written through these kernels.
It is easy to make a transformation from the simplest kernels to the distribution functions or to the coherency matrix.
Let us write the density matrix kernel for the system of $N$ photons and $N_{+}$ electrons
\beq \label{eq:kernelrhodef}
& \strut\disp \rho_{s_1...s_N,\sigma_1...\sigma_{N_{+}},n_1...n_{N_{+}}}^{s'_1
...s'_N,\sigma'_1...\sigma'_{N_{+}},n'_1...n'_{N_{+}}}\left({{\vk'_1...
\vk'_N}\atop {\vk_1...\vk_N}}\Biggl|{{Y'_1...Y'_{N_{+}}\;Z'_1...Z'_{N_{+}}}
\atop {Y_1...Y_{N_{+}}\;Z_1...Z_{N_{+}}}}\right)\equiv 
& \nonumber \\
& \strut\disp 
\equiv\left\langle c{^*}{_{n'_1...n'_{N_{+}},s'_1...s'_N}^{\sigma'_1...
\sigma'_{N_{+}}}}(Y'_1,Z'_1,...,Y'_{N_{+}}Z'_{N_{+}},\vk'_1,...,\vk'_N)
c_{n_1...n_{N_{+}},s_1...s_N}^{\sigma_1...\sigma_{N_{+}}}(Y_1,Z_1,...,
Y_{N_{+}}Z_{N_{+}},\vk_1,...,\vk_N)\right\rangle. &
\eeq
Further let us write some useful relations for the kernels. For the sake of simplicity we consider only the photon gas. 
These relations can be generalized trivially to the case of the electron-photon gas. 
The kernel for the system of $N$ photons can be written through the density matrix:
\be \label{eq:matrix2kernel}
\rho_{s_1...s_N}^{s'_1...s'_N}\left({\vk'_1...\vk'_N}
\atop {\vk_1...\vk_N}\right)=N!\ \ovl {\Psi}_{s_1...s_N}(\vk_1,...,\vk_N)
\rho_N\Psi_{s'_1...s'_N}(\vk'_1,...,\vk'_N).
\ee
Hereinafter we call it the $N$-particle kernel. It is normalized to unity:
\be \label{eq:normrhoN1}
\int\prod_{i=1}^N\frac{\d\vk_i}{k_i}\rho_{s_1...s_N}^{s_1...s_N}\left(
{\vk_1...\vk_N}\atop {\vk_1...\vk_N}\right)=\texttt {Sp} (\rho)=1.
\ee
For any $m\leq N$ the $m$-particle kernel can be calculated as 
\be \label{eq:rhomdef}
\rho_{s_1...s_m}^{s'_1...s'_m}\left({\vk'_1...\vk'_m}\atop {\vk_1...\vk_m}
\right)=\frac{1}{(N-m)!}\int\prod_{i=m+1}^N\frac{\d\vk_i}{k_i}\rho_{s_1...
s_m,s_{m+1}...s_N}^{s'_1...s'_m,s_{m+1}...s_N}\left({\vk'_1...\vk'_m \vk_{m+1}
...\vk_N}\atop {\vk_1...\vk_m \vk_{m+1}...\vk_N}\right).
\ee
The 1-particle kernel can be expressed through the $N$-particle kernel as 
\be \label{eq:rhosskk}
\rho_s^{s'}\left({\vk'}\atop {\vk}\right)=\frac{1}{(N-1)!}
\int\prod_{i=2}^N\frac{\d\vk_i}{k_i}\rho{_s^{s'}}{_{s_2...s_N}^{s_2...s_N}}\left(
{\vk'\;\vk_2...\vk_N}\atop {\vk\;\vk_2...\vk_N}\right). 
\ee
It is normalized to the total number of the particles:
\be \label{eq:normrho1}
\int\frac{\d^3k}{k}\rho_s^s\left({\vk}\atop {\vk}\right)=N.
\ee
The diagonal elements of 1-particle kernel compose the coherency matrix in the case of photon gas.

\subsubsection{Transformation from the simplest kernel to the distribution functions}

The transformation from 1-particle density matrix to the distribution function in the case of electrons 
in the case of field-free space can be made using the Wigner function. The Wigner function is defined as
\be \label{eq:rhovpvrdef}
\rho(\vp,\vr)=\int \d\vv\exp\left(i\vp\cdot\vv\right)\rho_s\left(
\vr+\vv/2,\vr-\vv/2\right),
\ee
where $\rho_s\left(\vr+\vv/2,\vr-\vv/2\right)$ is a 1-particle density matrix in the coordinate representation. The momentum and coordinate representations are connected through the Fourier transforms:    
\be \label{eq:rhosvrvr}
\rho_s(\vr,\vr\,')=\frac{1}{(2\pi)^6}\int\frac{\d\vp\ \d\vp'}
{\sqrt {p_0\ p'_0}}\exp\left(-i(\vp\cdot\vr-\vp\,'\cdot \vr\,')\right)
\rho\left({{\vp'}\atop {\vp}}\right).
\ee
Then one can rewrite the Wigner function using the density matrix in the momentum representation:
\be \label{eq:Wignerfunc}
\rho(\vp,\vr)=\frac{1}{(2\pi)^6}\int\frac{\d\vp_1\d\vp'_1}
{\sqrt {p_{01}p'_{01}}}\exp\left(-i(\vp_1-\vp_1{\!\!'})\cdot\vr\right)
\delta\left(\vp-\frac{\vp_1+\vp_1{\!\!'}}{2}\right)\rho\left(
{{\vp_1{\!\!'}}\atop {\vp_1}}\right).
\ee
The inverse transformation from the Wigner function to the density matrix in momentum representation reads
\be \label{eq:Wignerrev}
\rho\left({{\vp_1{\!\!'}}\atop {\vp_1}}\right)=\frac{\sqrt {p_{01}p'_{01}}}
{(2\pi)^3}\int\d\vp\ \d\vr\ \exp\left(i(\vp_1-\vp_1{\!\!'})\cdot\vr\right)
\delta\left(\vp-\frac{\vp_1+\vp_1{\!\!'}}{2}\right)\rho(\vp,\vr).
\ee
The time scale of the electron-photon interaction is much smaller than the time scale of noticeable changes of the distribution functions. Therefore, in the last equation one can assume that the Wigner function does not depend on the space variables. In this case, the integration could be made easily and we can write
\be \label{eq:nucl2df1}
\rho\left({{\vp_1{\!\!'}}\atop {\vp_1}}\right)=p_{01}\delta(\vp_1{\!\!'}-\vp_1)
\rho(\vp_1).
\ee
Then one can convert the 1-particle density matrix kernel to the distribution function in the case of spinless particles or to the coherency matrices in the case of particles with non-zero spin. In the last case a trivial generalization is used: 
\be \label{eq:nucl2df2}
\rho_{\tau_1}^{\tau'_1}\left({{\vp_1{\!\!'}}\atop {\vp_1}}\right)=p_{01}
\delta(\vp_1{\!\!'}-\vp_1)\rho_{\tau_1}^{\tau'_1}(\vp_1).
\ee
Equations (\ref{eq:nucl2df1}) and (\ref{eq:nucl2df2}) can be written immediately from the physical meaning of the 
density matrix and the assumption that the time scale of the interaction and the time between interactions are mush 
smaller than the time scale of noticeable changes of the distribution function. 

One can also write similar relation for the case of electrons in the external magnetic field.
If we consider a non-interacting electron in the B-field 
not accounting for  cyclotron radiation (which should be described by another kinetic equation), 
then the electron should conserve its $z$-momentum and the Landau level. 
This means that the kernel should be diagonal over both $Z$ and $n$, because it is not possible 
to have mixed states corresponding to different values of the $z$-projection of momentum or the Landau level. 
Non-diagonal elements in the kernel can appear only if one accounts for interactions between particles, 
but because of the smallness of the interaction time scale
the kernel should be diagonal over $z$-projection of momentum and the Landau levels.
In this case the relation will have the following form (a detailed derivation is given in Appendix \ref{Ritus}):
\be \label{eq:rhoRnZ}
\rho_{\sigma n}^{\sigma' n'}\left({{Y' Z'}\atop {Y Z}}\right)=
R_{n}(Z)\delta_{n}^{n'}\delta(Y'-Y)\delta(Z'-Z)\rho_{\sigma n}^{\sigma'}(Y,Z),
\ee
where $\sigma$ and $\sigma'$ describe the electron spin-states, 
$n$ and $n'$ are the Landau levels, $Z$ and $Z'$ are the momentum projections and 
$R_{n}(Z)$ is the electron energy  given by equation (\ref{eq:RnZ}).

A transformation from the 1-particle density matrix to the distribution function in momentum space is trivial, 
but one must again assume that the 
typical time scale of changes of the distribution function is much larger than the typical time scales of interaction between the particles.

\subsection{Description of the interaction}

\subsubsection{Description of the single interaction}

Let us mark parameters of the particles before the interaction with the subscript "i" and particles after interaction with the subscript "f". There are three conservation laws for Compton scattering in the magnetic field. They are the energy conservation, the conservation of the momentum along the magnetic field and the conservation of the transversal momentum:
\be \label{eq:conservlaws}
R_\ii+k_\ii=R_\ff+k_\ff,\quad 
Z_\ii+k_\ii\cos\theta_\ii=Z_\ff+k_\ff\cos\theta_\ff,\quad
Y_\ii+k_\ii\sin\theta_\ii\sin\varphi_\ii=Y_\ff+k_\ff\sin\theta_\ff\sin\varphi_\ff.
\ee
Let us use special designation for product of $\delta$-functions which are describing these conservation laws:
\beq& \strut\disp 
\delta\left(n,Y,Z,\vk\;|\;n',Y',Z',\vk'\right)\equiv
& \nonumber \\
& \strut\disp
\equiv\delta(R_n(Z)+k-R_{n'}(Z')-k')
\delta(Z+k\cos\theta-Z'-k'\cos\theta')
\delta(Y+k\sin\theta\sin\varphi-Y'- k'\sin\theta'\sin\varphi')&.
\eeq 
A single interaction can be described by the $S$-matrix. The elements of the $S$-matrix can be calculated using methods of quantum electrodynamics.
In the simplest case the elements of the $S$-matrix can be obtained using second-order perturbation theory. In this case Compton scattering can be represented by two Feynman diagrams with two vertices in both of them and one can write an expression for the $S$-matrix elements:
\be \label{eq:Sfi}
S_{\ff\,\ii}=-4\pi i\alpha\int\d^4r_1\d^4r_2\ovl {\Psi}_\ff(\ur_2)\left\{
\left[\ugam\uA^\dag_\ff(\ur_2)\right]G(\ur_2,\ur_1)\left[
\ugam\uA_\ii(\ur_1)\right]+\left[\ugam\uA_\ii(\ur_2)\right]G(\ur_2,\ur_1)
\left[\ugam\uA^\dag_\ff(\ur_1)\right]\right\}\Psi_\ii(\ur_1),
\ee
where $\ugam\uA$ is the Dirac inner product of a 4-vector and $\ugam$-matrix, and $G(\ur_2,\ur_1)$ is a relativistic electronic propagator in the presence of a constant magnetic field, $\Psi_\ii(\ur)$ and $\Psi_\ff(\ur)$ are the electron wave-functions in coordinate representation, and 
$\alpha=e^2$  is the fine-structure constant.   
The $S$-matrix elements and the cross-sections for Compton scattering in magnetic field contain resonances which have to be 
regularized \cite{NK1993}. The calculations are not trivial and have been performed  only in special cases \cite{Her1979,DH1986,BAM1986,GHBCM2000}.

\subsubsection{Evolution of the density matrix}

The evolution of the density matrix can be described by equation:
\be \label{eq:drhodt}
i\Dr {\rho(t)}{t}=H(t)\rho(t)-\rho(t)H(t),
\ee
where the Hamiltonian is
\be \label{eq:Hamiltonian}
H(t)=-e\int\d\vr\ \ovl {\psi}(\ur)\ugam\uA(\ur)\psi(\ur).
\ee
Equation (\ref{eq:drhodt}) is written here in non-covariant form, but it can be transformed to the explicitly covariant form using the Tomonaga-Schwinger equation \cite{BS1959}. It means that the form of the equation is covariant for the longitudinal Lorentz transformations (along the magnetic field direction). 
The solution of equation (\ref{eq:drhodt}) can be presented by the operator of evolution $U(x,y)$:
\be \label{eq:drhodtsol}
\intl_{t_0}^t\Dr {\rho(t')}{t'}\d t'=\rho(t)-\rho(t_{0})=U(t,t_0)\rho(t_0)-
\rho(t_0)U(t,t_0).
\ee
On the other hand the operator $U(t,t_{0})$ can be represented in the following form:
\be \label{eq:Utt0}
U(t,t_0)=\intl_{t_0}^t\d t'\intl_{t_0}^t\d t''\intl_V\d\vr'
\intl_V\d\vr''\cS(\ur',\ur'')=\intl_\cV\d^4r'
\intl_{\cV}\d^{4}r''\cS(\ur',\ur''),
\ee
where $\cV=[t_0,t]\times V$ is the volume in Minkowski space and
\be \label{eq:cSurur}
\cS(\ur',\ur'')=i\frac{e^{2}}{(2\pi)^{9}}\int\frac {\d Y\d Z}{R}\frac {\d Y'\d Z'}{R'}
\frac {\d\vk}{k}\frac {\d\vk'}{k'}b^\dagger_{n'\sigma'}(Z')b_{n\sigma}(Z)
\bar {a}_{(s')}(\vk')a_{(s)}(\vk)\cN{_n^{n'}}{_\sigma^{\sigma'}}{_s^{s'}}
\left(\left.{{\vk'}\atop{\vk}}{{Y'}\atop{Y}}{{Z'}\atop{Z}}\right|{{\ur'}\atop {\ur''}}\right).
\ee
The space integral of $\cN$ can be represented through the elements of the scattering $M$-matrix:
\be \label{eq:intd4rcN}
\int\limits_{\cV}\d^4r'\int\limits_{\cV}\d^4r''\cN{_n^{n'}}{_\sigma^{\sigma'}}{_s^{s'}}
\left(\left.{{\vk'}\atop{\vk}}{{Y'}\atop{Y}}{{Z'}\atop{Z}}\right|{{\ur'}\atop {\ur''}}\right)=
(2\pi)^{8}\delta\left(n,Y,Z,\vk\;|\;n',Y',Z',\vk'\right)
M^{\sigma's'}_{\sigma s}\left(\left.{{n'Y'Z'}\atop {n Y Z}}\right|{{\vk'} \atop
{\vk}}\right),
\ee
then  $U(t,t_0)$ can be rewritten in the following form
\be \label{eq:UfM}
U(t,t_0)=\frac{i}{2\pi}\alpha 
\int\frac {\d Y\d Z}{R}\frac {\d Y'\d Z'}{R'}
\frac {\d\vk}{k}\frac {\d\vk'}{k'}b^\dagger_{n'\sigma'}(Z')b_{n\sigma}(Z)
\bar {a}_{(s')}(\vk')a_{(s)}(\vk)M^{\sigma's'}_{\sigma s}
\left(\left.{{n'Y'Z'}\atop {n Y Z}}\right|{{\vk'}\atop {\vk}}\right).
\ee
Let us assume that the typical time scale of the density matrix changes is much larger than the typical time scales of a single interaction. 
In that case changes in the distribution on a macroscopically small times scale can be represented througth the $S$-matrix because
 $M$-matrix can be considered as the scattering $S$-matrix divided by the fine-structure constant:
\be
M^{\sigma_\ff s_\ff}_{\sigma_\ii s_\ii}\left(\left.{{n_\ff Y_\ff Z_\ff}\atop {n_\ii Y_\ii Z_\ii}}\right|{{\vk_\ff} \atop{\vk_\ii}}\right)\equiv
M_{\rm fi}=\frac{S_{\rm fi}}{\alpha},
\ee 
and the time interval $[t_0, t_0 +t]$ is considered as a macroscopically small time.
Equations (\ref{eq:drhodtsol}) and (\ref{eq:UfM}) determine the solution formulated through the elements of the scattering matrix.  
We reformulate these equations below in terms of the kernels of the density matrix.

\section{Derivation of the kinetic equation for the photon gas}

\subsection{Methodology of the kinetic equation derivation}

\subsubsection{Summary of our assumptions and the Bogolyubov method}

We derive kinetic equation using a generalization of the Bogolyubov method (for the case of quantum statistics). At the first step we formulate  Liouville's theorem in terms of the kernels of density matrix. One can derive the equations for kernels of different orders (1-particle, 2-particle and other) by integrating over the parameters of different numbers of particles. We use this method to obtain the system of kinetic equations. If the full ensemble contains $N$ particles, the system of equations contains $N$ equations. 
In the case of the rarefied gas one can use only a few first equations from the Bogolyubov hierarchy. The criterion of rarefaction can be formulated 
through the ``gaseous parameter'', which depends on the concentration of the particles and the cross sections of their interaction:
\be \label{eq:alphagas}
\alpha_{\rm gas} \equiv 
\sqrt {\sigma_{\rm T}}(n_{\rm e}n_{\ph})^{1/6} \ll 1, 
\ee
where $\sigma_{\rm T}$ is the Thomson cross section, $n_{\rm e}$ and $n_\ph$ are the electron and the photon concentrations, correspondingly. 
According to the principle of weakening of correlations, which is satisfied for sufficiently rarefied gases, the correlations are accounted for only in the equation for the 1-particle matrix via kernel by entering the right-hand side ({rhs}) of the aforementioned equation. This kernel is assumed to characterize the electron and photon states after the interaction. It can be represented via the same kernel before the interaction and the correlation function. 
To derive the kinetic equation for the typical conditions in the neutron star atmospheres, it is enough to use only the first and the second equation from the Bogolyubov hierarchy.

\subsubsection{Formulation of  Liouville's theorem and the equations of Bogolyubov hierarchy}

We use the following notations: $R\equiv R_n(Z),\quad R'\equiv R_{n'}(Z')$, etc.
There are $N$ photons and $N_+$ electrons in the system. The equation, describing the change of $(N+N_{+})$-particle kernel during macroscopically small time $T_{0}$ is written as
\beq \label{eq:comptLiuvTh}
& \strut\disp \rho_{s_1...s_N \sigma_1...\sigma_{N_{+}} n_1...n_{N_{+}}}
^{s'_1..s'_N \sigma'_1...\sigma'_{N_{+}} n'_1...n'_{N_{+}}}
\left({{\vk'_1...\vk'_N}\atop {\vk_1...\vk_N}}\left|{{Y'_1...Y'_{N_{+}} Z'_1 
... Z'_{N_{+}}}\atop {Y_1...Y_{N_{+}} Z_1...Z_{N_{+}}}}\right|
\frac {T_0}{2}\right) 
& \nonumber \\
& \strut\disp 
=\rho_{s_1..s_N \sigma_1...\sigma_{N_{+}} n_1...n_{N_{+}}}^{s'_1
...s'_N \sigma'_1...\sigma'_{N_{+}} n'_1...n'_{N_{+}}}\left({{\vk'_1...\vk'_N}
\atop {\vk_1...\vk_N}}\left|{{Y'_1...Y'_{N_{+}} Z'_1...Z'_{N_{+}}}
\atop {Y_1...Y_{N_{+}} Z_1...Z_{N_{+}}}}\right|-\frac {T_0}{2}\right)+i
\frac {\alpha}{2\pi}\int\frac {\d Y\d Z}{R}
\frac {\d Y'\d Z'}{R'}\frac {\d\vk}{k}\frac {\d\vk'}{k'}  
& \nonumber \\
& \strut\disp 
\times\delta\left(n,Y,Z,\vk\;|\;n',Y',Z',\vk'\right)\ 
b^\dag_{n'\sigma'}(Y',Z')b_{n\sigma}(Y,Z)\bar {a}_{s'}(\vk')a_s(\vk)
M^{\sigma's'}_{\sigma s}\left({{n'}\atop{n}}{{Y'}\atop{Y}}{{Z'}\atop{Z}}\left|{{\vk'}
\atop {\vk}}\right) \right. 
& \nonumber \\
& \strut\disp 
\times\sum^N_{i=1}\sum^{N_{+}}_{i_{+}=1}\left[
\delta^{s'}_{s_i'}\delta(\vk'-\vk'_i)\delta^{n'}_{n'_{i_{+}}}
\delta^{\sigma'}_{\sigma'_{i_{+}}}\delta(Y'-Y'_{i_{+}})\delta(Z'-Z'_{i_{+}})
 \right. & \nonumber \\
& \strut\disp \times\rho
_{s_1...s_i...s_N \sigma_1...\sigma_{i_{+}}...\sigma_{N_{+}} n_1...n_{i_{+}}...n_{N_{+}}}
^{s'_1...s...s'_N \sigma'_1...\sigma...\sigma'_{N_{+}} n'_1...n...n'_{N_{+}}}
\left({{\vk'_1...\vk...
\vk'_N}\atop {\vk_1...\vk_i...\vk_N}}\right|{{Y'_1...Y...Y'_{N_{+}} Z'_1...
Z...Z'_{N_{+}}}\atop {Y_1...Y_{i_{+}}...Y_{N_{+}} Z_1...Z_{i_{+}}...
Z_{N_{+}}}}\left|-\frac {T_0}{2}\right) 
& \nonumber \\
& \strut\disp 
-\delta^s_{s_i}\delta(\vk-\vk_i)\delta^n_{n_{i_{+}}}
\delta^\sigma_{\sigma_{i_{+}}}\delta(Y-Y_{i_{+}})\delta(Z-Z_{i_{+}}) 
 & \nonumber \\
& \strut\disp 
\times\rho_{s_1...s'...s_N \sigma_1...\sigma'...\sigma_{N_{+}} 
n_1...n'...n_{N_{+}}}^{s'_1...s'_i...s'_N \sigma'_1...\sigma'_{i_{+}}...
\sigma'_{N_{+}} n'_1...n'_{i_{+}}...n'_{N_{+}}}\left(\left.{{\vk'_1...\vk'_{i}
...\vk'_N}\atop {\vk_1...\vk'...\vk_N}}\right|{{Y'_1...Y'_{i_{+}}...
Y'_{N_{+}} Z'_1...Z'_{i_{+}}...Z'_{N_{+}}}\atop {Y_1...Y'...Y_{N_{+}} Z_1 
...Z'... Z_{N_{+}}}}\left|-\frac {T_0}{2}\right)\right]. &
\eeq
This equation is a formulation of  Liouville's theorem in terms of the density matrix kernels.
It can be used to find the equation describing the evolution of 1-particle photon kernels on the time interval $\left[-T_0/2,T_0/2\right]$, 
which is the first equation of the Bogolyubov hierarchy. 

We integrate equation (\ref{eq:comptLiuvTh}) over the parameters of $(N-1)$ photons and $N_{+}$ electrons:
\beq
& \strut\disp \rho_{s_1}^{s'_1}\left({{\vk'_1}\atop {\vk_1}}\left|
\frac{T_0}{2}\right)-\rho_{s_1}^{s'_1}\left({{\vk'_1}\atop {\vk_1}}\right|
-\frac{T_0}{2}\right)=\frac {1}{(N-1)!N_{+}!}i
\frac{\alpha}{2\pi}\int\frac {\d Y\d Z}{R}\frac {\d Y'\d Z'}{R'}\frac {\d\vk}{k}
\frac {\d\vk'}{k'}\delta\left(n,Y,Z,\vk\;|\;n',Y',Z',\vk'\right) 
&\\
& \strut\disp 
\times b^\dag_{n'\sigma'}(Y',Z')b_{n\sigma}(Y,Z)\bar {a}_{s'}
(\vk')a_s(\vk)
M^{\sigma's'}_{\sigma s}\left({{n'}\atop{n}}{{Y'}\atop{Y}}{{Z'}\atop{Z}}\right.\left|{{\vk'}\atop {\vk}}\right)
\sum_{i_{+}=1}^{N_{+}}\int\prod_{i=2}^N
\left(\frac {\d\vk_i}{k_i}\frac {\d\vk'_i}{k'_i}\right)\prod_{i_{+}=1}^{N_{+}}
\left(\frac {\d Y_{i_{+}}\d Z_{i_{+}}}{R_{i_{+}}}\frac{\d Y'_{i_{+}}\d Z'_{i_{+}}}{R'_{i_{+}}}
\right) 
& \nonumber \\
& \strut\disp 
\times \left\{
\delta^{n'}_{n'_{i_{+}}}\delta^{\sigma'}_{\sigma'_{i_{+}}}\delta(Y'-Y'_{i_{+}})
\delta(Z'-Z'_{i_{+}})\right. 
\left[\delta ^{s'}_{s_{1}'}\delta(\vk'-\vk'_{1})
P^{s}_{s_1}\left(\left.{{\vk}\atop{\vk_1}}\right|1,1,\textit{J},\textit{G}\right)\right. 
\left.+\sum_{i=2}^{N}\delta^{s'}_{s_i}\delta(\vk'-\vk_i)
P^{s'_1}_{s_1}\left(\left.{{\vk'_1}\atop{\vk_1}}\right|i,i_{+},\textit{J},\textit{G}\right)
\right] 
 & \nonumber \\
& \strut\disp -\delta^n_{n_{i_{+}}}\delta^\sigma_{\sigma_{i_{+}}}
\delta(Y-Y_{i_{+}})\delta(Z-Z_{i_{+}})  
\left[\delta ^s_{s_1}\delta(\vk-\vk_1)
R^{s'_1}_{s'}\left(\left.{{\vk'_1}\atop{\vk'}}\right|1,1,\textit{J'},\textit{G'}\right)
\right. 
\left.\left.+\sum_{i=2}^{N}\delta ^{s}_{s_{i}}\delta(\vk-\vk_{i})
R^{s'_1}_{s_1}\left(\left.{{\vk'_1}\atop{\vk_1}}\right|i,i_{+},\textit{J'},\textit{G'}\right)
\right]
\right\} , &\nonumber
\eeq
where the kernels under the integral in the rhs of the equation correspond to time $t=-T_0/2$ and where
we used special designations for kernels describing the system of $N$ photons and $N_+$ electrons:
$$P^{s_a}_{s_b}\left(\left.{{\vk_c}\atop{\vk_d}}\right|i,j,\textit{J},\textit{G}\right)\equiv \rho{^{s_{a}s_{2}...}_{s_{b}s_{2}...}}{^{s ...}_{s_i ...}}^{s_N}_{s_N}
{^{\sigma_{1}\sigma_{2}...\sigma...\sigma_{N+}}_{\sigma_{1}\sigma_{2}...\sigma_j...\sigma_{N+}}}
{^{n_{1}n_{2}...n...n_{N+}}_{n_{1}n_{2}...n_j...n_{N+}}}
\left({{\vk_c \vk_2 ...}\atop{\vk_d \vk_2 ...}}
{{\vk...}\atop{\vk_i ...}}{{\vk_N}\atop{\vk_N}}\left|
{{Y_1 Y_2 ... Y ... Y_{N_{+}}}\atop{Y_1 Y_2 ... Y_{j} ... Y_{N_{+}}}}
{{Z_1 Z_2 ... Z ... Z_{N_{+}}}\atop{Z_1 Z_2 ... Z_{j} ... Z_{N_{+}}}}
\right)\right.,
$$
$$R^{s_a}_{s_b}\left(\left.{{\vk_c}\atop{\vk_d}}\right|i,j,\textit{J},\textit{G}\right)\equiv \rho{^{s_{a}s_{2}...}_{s_{b}s_{2}...}}{^{s_i ...}_{s ...}}^{s_N}_{s_N}
{^{\sigma_{1}\sigma_{2}...\sigma_j...\sigma_{N+}}_{\sigma_{1}\sigma_{2}...\sigma...\sigma_{N+}}}
{^{n_{1}n_{2}...n_j...n_{N+}}_{n_{1}n_{2}...n...n_{N+}}}
\left({{\vk_c \vk_2 ...}\atop{\vk_d \vk_2 ...}}
{{\vk_i ...}\atop{\vk ...}}{{\vk_N}\atop{\vk_N}}\left|
{{Y_1 Y_2 ... Y_j ... Y_{N_{+}}}\atop{Y_1 Y_2 ... Y ... Y_{N_{+}}}}
{{Z_1 Z_2 ... Z_j ... Z_{N_{+}}}\atop{Z_1 Z_2 ... Z ... Z_{N_{+}}}}
\right)\right.,
$$
where $\textit{J}=(s,\vk)$ and $\textit{G}=(\sigma,n,Y,Z)$ are parameters of photons and electrons respectively.
Let us denote the terms under the sum $\sum\limits_{i=2}^{N}$ by $\Xi$. One can transform it:
\beq
& \strut\disp \Xi=\frac{1}{(N-1)!N_{+}!}i
\frac {\alpha}{2\pi}
\int\frac {\d Y\d Z}{R}\frac {\d Y'\d Z'}{R'}\frac {\d\vk}{k}\frac {\d\vk'}{k'}
\delta\left(n,Y,Z,\vk\;|\;n',Y',Z',\vk'\right)  
& \nonumber \\
& \strut\disp 
\times b^\dag_{n'\sigma'}(Y',Z')b_{n\sigma}(Y,Z)\bar {a}_{s'}
(\vk')a_s(\vk) 
M^{\sigma's'}_{\sigma s}\left({{n'}\atop{n}}{{Y'}\atop{Y}}{{Z'}\atop{Z}}\right.\left|{{\vk'}\atop {\vk}}\right)
\sum_{i_{+}=1}^{N_{+}}\int\prod_{i=2}^{N}
\left(\frac {\d\vk_i}{k_i}\frac {\d\vk'_i}{k'_i}\right)\prod_{i_{+}=1}^{N_{+}}
\left(\frac{\d Y_{i_{+}}\d Z_{i_{+}}}{R_{i_{+}}}\frac{\d Y'_{i_{+}}\d Z'_{i_{+}}}{R'_{i_{+}}}\right)  
& \nonumber \\
& \strut\disp \times\left\{
\delta^{n'}_{n'_{i_{+}}}\delta^{\sigma'}_{\sigma'_{i_{+}}}\delta(Y'-Y'_{i_{+}})
\delta(Z'-Z'_{i_{+}})  \right. 
\sum_{i=2}^N\delta ^{s'}_{s_i}\delta(\vk'-\vk_i)
P^{s'_1}_{s_1}\left(\left.{{\vk'_1}\atop{\vk_1}}\right|i,i_{+},\textit{J},\textit{G}\right) 
& \nonumber \\
& \strut\disp 
-\delta^n_{n_{i_{+}}}\delta^{\sigma}_{\sigma_{i_{+}}}
\delta(Y-Y_{i_{+}})\delta(Z-Z_{i_{+}})  
\left.\sum_{i=2}^{N}\delta^s_{s_{i}}\delta(\vk-\vk_i)
R^{s'_1}_{s_1}\left(\left.{{\vk'_1}\atop{\vk_1}}\right|i,i_{+},\textit{J'},\textit{G'}\right)
\right\} 
& \nonumber \\
& \strut\disp 
=\frac {1}{(N-1)!N_{+}!}i 
\frac {\alpha}{2\pi}
\int\d Y\d Z\d Y'\d Z'\frac {\d\vk}{k}\frac {\d\vk'}{k'}\delta\left(n,Y,Z,\vk\;|\;n',Y',Z',\vk'\right) 
& \nonumber \\
& \strut\disp 
\times\sum\limits_{i_{+}=1}^{N_{+}}\int\prod_{i=2}^{N}\left(\frac {\d\vk_i}{k_i}
\frac {\d\vk'_i}{k'_i}\right)\prod_{i_{+}=1}^{N_{+}}\left(\frac {\d Y_{i_{+}}
\d Z_{i_{+}}}{R_{i_{+}}}\frac {\d Y'_{i_{+}}\d Z'_{i_{+}}}{R'_{i_{+}}}\right)  
\left\{
\sum_{i=2}^{N}
P^{s'_1}_{s_1}\left(\left.{{\vk'_1}\atop{\vk_1}}\right|i,i_{+},\textit{J'},\textit{G'}\right) 
\right. 
\left. -\sum_{i=2}^{N}
P^{s'_1}_{s_1}\left(\left.{{\vk'_1}\atop{\vk_1}}\right|i,i_{+},\textit{J'},\textit{G'}\right)\right\}. 
& \nonumber
\eeq
Creation and annihilation operators with the elements of scattering matrix were placed under the integral in the last transformation. Indices in the round
brackets indicate the positions of pairs of parameters ($s$--$s'$), ($n$--$n'$), ($Y$--$Y'$), ($Z$--$Z'$), ($\sigma$--$\sigma'$) and ($\vk$--$\vk$'). One notices that all the terms in the sum $\Xi$ cancel out, giving $\Xi=0$.
The remaining part of the equation can be rewritten as
\beq \label{eq:ker1}
& \strut\disp 
\rho_{s_1}^{s'_1}\left({{\vk'_1}\atop {\vk_1}}\right.\left|
\frac {T_0}{2}\right)-\rho_{s_1}^{s'_1}\left({{\vk'_1}\atop {\vk_1}}\left|-
\frac {T_0}{2}\right)=i
\frac {\alpha}{2\pi}\int
\frac {\d Y\d Z}{R}\frac {\d Y'\d Z'}{R'}\frac {\d\vk}{k}\frac {\d\vk'}{k'}
\delta\left(n,Y,Z,\vk\;|\;n',Y',Z',\vk'\right)\right.
& \nonumber \\
& \strut\disp 
\times M^{\sigma's'}_{\sigma s}\left({{n'}\atop{n}}{{Y'}\atop{Y}}{{Z'}\atop{Z}}\right.\left|{{\vk'}\atop {\vk}}\right) 
\left[\delta^{s'}_{s'_1}\delta(\vk'-\vk'_1)
\rho_{s_1 \sigma' n'}^{s \sigma n}\left(\left.{{\vk}\atop {\vk_1}}\right|
{{Y Z}\atop {Y' Z'}}\right)-\delta^s_{s_1}\delta(\vk-\vk_1)
\rho_{s' \sigma' n'}^{s'_1 \sigma n}\left(\left.{{\vk'_1}\atop {\vk'}}\right|
{{Y Z}\atop {Y' Z'}}\right)\right]. &
\eeq
Thus, we have obtained the first equation of the Bogolyubov hierarchy describing the evolution of 1-particle density matrix kernel through the 2-particle density matrix kernel.  

Let us now obtain the second equation of  the Bogolyubov hierarchy  for the 2-particle kernels.  We proceed with the integration and the summation over the parameters of $(N-1)$ photons and $(N_{+}-1)$ electrons in equation (\ref{eq:comptLiuvTh}):
\beq
& \strut\disp \rho_{s_1 \sigma_1 n_1}^{s'_1 \sigma'_1 n'_1}\left({{\vk'_1 Y'_1 
Z'_1}\atop {\vk_1 Y_1 Z_1}}\left|\frac {T_0}{2}\right)-
\rho_{s_1 \sigma_1 n_1}^{s'_1 \sigma'_1 n'_1}\left({{\vk'_1 Y'_1 Z'_1}\atop
{\vk_1 Y_1 Z_1}}\right|-\frac {T_0}{2}\right) 
=\frac {1}{(N-1)!(N_{+}-1)!} 
i\frac {\alpha}{2\pi} 
& \nonumber \\
& \strut\disp
\times\int\frac {\d Y\d Z}{R}\frac {\d Y'\d Z'}{R'}\frac {\d\vk}{k}\frac {\d\vk'}{k'}
\delta\left(n,Y,Z,\vk\;|\;n',Y',Z',\vk'\right)b^\dag_{n'\sigma'}(Y',Z')b_{n\sigma}(Y,Z)\bar {a}_{s'}(\vk')a_{s}(\vk)
M^{\sigma's'}_{\sigma s}\left({{n'}\atop{n}}{{Y'}\atop{Y}}{{Z'}\atop{Z}}\right.\left|{{\vk'}\atop {\vk}}\right)  & \nonumber \\
& \strut\disp \times\int\prod_{i=2}^{N}\left(\frac {\d\vk_i}{k_i}
\frac {\d\vk'_i}{k'_i}\right)\prod_{i_{+}=2}^{N_{+}}\left(\frac {\d Y_{i_{+}}
\d Z_{i_{+}}}{R_{i_{+}}}\frac {\d Y'_{i_{+}}\d Z'_{i_{+}}}{R'_{i_{+}}}\right)
 \times\! 
\left\{\!
\left[\delta^{n'}_{n'_1}\delta^{\sigma'}_{\sigma'_1}\delta^{s'}_{s'_1}
\delta(Y'\!\!-\!Y'_1)\delta(Z'\!\!-\!Z'_1)\delta(\vk'\!\!-\!\vk'_1)\ 
P^{s}_{s_1}\left(\left.{{\vk}\atop{\vk_1}}\right|1,1,\textit{J},\textit{G}\right)
\right.\right. & \nonumber \\
& \strut\disp 
+\delta^{s'}_{s'_1}\delta(\vk'-\vk'_1)\sum_{i_{+}=2}^{N_{+}}
\delta_{\sigma_{i_+}}^{\sigma'}\delta_{n_{i_+}}^{n'}\delta(Y_{i_+}-Y')
\delta(Z_{i_+}-Z')\   
P^{s}_{s_1}\left(\left.{{\vk}\atop{\vk_1}}\right|1,i_{+},\textit{J},\textit{G}\right)
 & \nonumber \\
& \strut\disp +\delta^{n'}_{n'_1}\delta^{\sigma'}_{\sigma'_{1}}\delta(Y'-Y'_1)
\delta(Z'-Z'_1)\sum_{i=2}^N\delta_{s_i}^{s'}\delta(\vk-\vk_i)\  
P^{s_1}_{s_1}\left(\left.{{\vk_1}\atop{\vk_1}}\right|i,1,\textit{J},\textit{G}\right)
& \nonumber \\
& \strut\disp +\sum_{i=2}^{N}\sum_{i_{+}=2}^{N_{+}}
\delta_{\sigma_{i_+}}^{\sigma'}\delta_{n_{i_+}}^{n'}\delta(Y_{i_+}-Y')
\delta(Z_{i_+}-Z')\delta_{s_i}^{s'}\delta(\vk-\vk_i)  
\left.\ 
P^{s_1}_{s_1}\left(\left.{{\vk_1}\atop{\vk_1}}\right|i,i_{+},\textit{J},\textit{G}\right)
\right]  & \nonumber \\
& \strut\disp -\left[\delta_s^{s_1}\delta_n^{n_1}\delta_{\sigma}^{\sigma_1}
\delta(\vk\!-\!\vk_1)\delta(Y\!-\!Y_1)\delta(Z\!-\!Z_1)\ 
R^{s_1}_{s'}\left(\left.{{\vk_1}\atop{\vk'}}\right|1,1,\textit{J'},\textit{G'}\right)\right.
& \nonumber \\
& \strut\disp +\delta_s^{s_1}\delta(\vk-\vk_1)\sum_{i_{+}=2}^{N_{+}}
\delta_{n_{i_+}}^n\delta_{\sigma_{i_+}}^{\sigma}\delta(Y-Y_{i_+})
\delta(Z-Z_{i_+}) \  
R^{s_1}_{s'}\left(\left.{{\vk_1}\atop{\vk'}}\right|1,i_{+},\textit{J'},\textit{G'}\right) 
 & \nonumber \\
& \strut\disp +\delta_n^{n_1}\delta_\sigma^{\sigma_1}\delta(Y-Y_1)\delta(Z-Z_1)
\sum_{i=2}^N\delta_{s_i}^s\delta(\vk-\vk_i)  \ 
R^{s_1}_{s_1}\left(\left.{{\vk_1}\atop{\vk_1}}\right|i,1,\textit{J'},\textit{G'}\right)
& \nonumber \\
& \strut\disp +\sum_{i=2}^N\sum_{i_{+}=2}^{N_{+}}\delta_{s_i}^s
\delta_{n_{i_+}}^n\delta_{\sigma_{i_+}}^\sigma\delta(Y-Y_{i_+})
\delta(Z-Z_{i_+})\delta(\vk-\vk_i)   
\left.\left.\ 
R^{s_1}_{s_1}\left(\left.{{\vk_1}\atop{\vk_1}}\right|i,i_{+},\textit{J'},\textit{G'}\right)
\right]\right\}. &
\eeq
The terms with a double sum $\disp\sum_{i=2}^N\sum_{i_{+}=2}^{N_{+}}$ cancel each other. 
Other terms are transformed into the form containing 2- and 3-particle kernels. 
As a result, we obtain the equation for the 2-particle density matrix kernel:
\beq \label{eq:ker2}
& \strut\disp \rho_{s_1 \sigma_1 n_1}^{s'_1 \sigma'_1 n'_1}\left(\left.
{{\vk'_1}\atop {\vk_1}}\right|{{Y'_1 Z'_1}\atop {Y_1 Z_1}}\left|
\frac {T_0}{2}\right)-\rho_{s_1 \sigma_1 n_1}^{s'_1 \sigma'_1 n'_1}\left(\left.
{{\vk'_1}\atop {\vk_1}}\right|{{Y'_1 Z'_1}\atop {Y_1 Z_1}}\right|-
\frac {T_0}{2}\right) 
& \nonumber \\
& \strut\disp 
=i\frac{\alpha}{2\pi}\int\frac {\d Y\d Z}{R}
\frac {\d Y'\d Z'}{R'}\frac {\d\vk}{k}\frac {\d\vk'}{k'}\delta\left(n,Y,Z,\vk\;|\;n',Y',Z',\vk'\right)
M^{\sigma's'}_{\sigma s}\left({{n'}\atop{n}}{{Y'}\atop{Y}}{{Z'}\atop{Z}}\right.\left|{{\vk'}\atop {\vk}}\right)  & \nonumber \\
& \strut\disp \times\left[\delta_{s'_1}^{s'}\delta_{n'_1}^{n'}
\delta_{\sigma'_1}^{\sigma'}\delta(Y'-Y'_1)\delta(Z'-Z'_1)\delta(\vk'-\vk'_1)\ 
\rho_{s_1 \sigma_1 n_1}^{s \sigma n}\left(\left.{{\vk}\atop {\vk_1}}\right|
{{Y Z}\atop {Y_1 Z_1}}\right)\right.  & \nonumber \\
& \strut\disp 
-\delta_{s_1}^s\delta_{n_1}^n\delta_{\sigma_1}^\sigma
\delta(Y-Y_1)\delta(Z-Z_1)\delta(\vk-\vk_1)\ \rho_{s' \sigma' n'}^{s'_1 
\sigma'_1 n'_1}\left(\left.{{\vk'_1}\atop {\vk'}}\right|{{Y'_1 Z'_1}\atop
{Y' Z'}}\right)  
& \nonumber \\
& \strut\disp 
+\delta_{n'_1}^{n'}\delta_{\sigma'_1}^{\sigma'}
\delta(Y'\!-\!Y'_1)\delta(Z'\!-\!Z'_1)\ \rho_{s_1 s' \sigma_1 n_1}^{s'_1 s 
\sigma n}\left(\left.{{\vk'_1 \vk}\atop {\vk_1 \vk'}}\right|{{Y Z}\atop
{Y_1 Z_1}}\right)\!-\!\delta_{n_1}^n\delta_{\sigma_1}^\sigma\delta(Y\!-\!Y_1)
\delta(Z\!-\!Z_1)\ \rho_{s_1 s' \sigma'_1 n'_1}^{s'_1 s \sigma' n'}\left(\left.
{{\vk'_1 \vk}\atop {\vk_1 \vk'}}\right|{{Y' Z'}\atop {Y'_1 Z'_1}}\right) 
& \nonumber \\
& \strut\disp 
\left. +\delta_{s'_1}^{s'}\delta(\vk'-\vk'_1)\ \rho_{s_1 \sigma_1 
n_1 \sigma' n'}^{s \sigma'_1 n'_1 \sigma n}\left(\left.{{\vk}\atop
{\vk_1}}\right|{{Y'_1 Z'_1 Y Z}\atop {Y_1 Z_1 Y' Z'}}\right)-
\delta_{s_1}^s\delta(\vk-\vk_1)\ \rho_{s' \sigma_1 n_1 \sigma' n'}^{s'_1 
\sigma'_1 n'_1 \sigma n}\left(\left.{{\vk'_1}\atop {\vk'}}\right|
{{Y'_1 Z'_1 Y Z}\atop {Y_1 Z_1 Y' Z'}}\right)\right]. &
\eeq
Thus, we have derived the equations for kernels of 1-particle density matrix of photons (\ref{eq:ker1}) and for the 
2-particle density matrix (\ref{eq:ker2}), which includes both the photon and the electron parameters.

\subsection{Completion of the derivation}

\subsubsection{Expression for the 2-particle kernels through the 1-particle kernels}

Equations (\ref{eq:ker1}) and (\ref{eq:ker2}) are the first two equations of the Bogolyubov hierarchy. The hierarchy can be continued, but using the principle of weakening of correlations, we have stopped at the first two equations. To obtain an equation describing the evolution of the 1-particle kernel through the 1-particle kernels, one must use these two equations. We  use the ``molecular chaos'' approximation, according to which there is no correlation between the distribution functions of photons and electrons before interaction. This approximation works better in the case, when the typical time between the interactions is much larger than the typical time of an interaction. The independence of the photons and electrons distributions can be expressed through the following equation:
\be \label{eq:m_chaos}
\rho_{\sigma n s}^{\sigma'n's'}\left({{\vk'}\atop {\vk}}\left|{{Y'Z'}\atop {Y Z}}
\right)=\rho_{\sigma n}^{\sigma'n'}\left({Y'Z'}\atop {Y Z}\right)\rho_{s}^{s'}
\left({{\vk'}\atop {\vk}}\right). \right.
\ee
The 2-particle kernels of the photon gas and the 2-particle kernels of the electron gas are presented through 1-particle kernels according to the properties of symmetry and anti-symmetry of bosonic and fermionic wave functions:
\beq \label{eq:exch_f}
& \strut\disp \rho_{s_1s_2}^{s'_1s'_2}\left({{\vk'_1\vk'_2}\atop {\vk_1\vk_2}}
\right)=\rho_{s_1}^{s'_1}\left({{\vk'_1}\atop {\vk_1}}\right)\rho_{s_2}^{s'_2}
\left({{\vk'_2}\atop {\vk_2}}\right)+\rho_{s_1}^{s'_2}\left({{\vk'_2}\atop
{\vk_1}}\right)\rho_{s_2}^{s'_1}\left({{\vk'_1}\atop {\vk_2}}\right),
 & \\ \label{eq:exch_e}
& \strut\disp \rho_{\sigma_1n_1\sigma_2n_2}^{\sigma'_1n'_1\sigma'_2n'_2}
\left({{Y'_1 Z'_1 Y'_2 Z'_2}\atop {Y_1 Z_1 Y_2 Z_2}}\right)=\rho_{\sigma_1n_1}^{\sigma'_1n'_1}
\left({{Y'_1 Z'_1}\atop {Y_1 Z_1}}\right)\rho_{\sigma_2n_2}^{\sigma'_2n'_2}
\left({{Y'_2 Z'_2}\atop {Y_2 Z_2}}\right)-\rho_{\sigma_1n_1}^{\sigma'_2n'_2}
\left({{Y'_2 Z'_2}\atop {Y_1 Z_1}}\right)\rho_{\sigma_2n_2}^{\sigma'_1n'_1}
\left({{Y'_1 Z'_1}\atop {Y_2 Z_2}}\right). &
\eeq
These equations become more accurate, when photons and electrons gases are sufficiently rarefied.
Transformations of the 3-particle kernels in equations (\ref{eq:ker1}) and (\ref{eq:ker2}) are simple because there are no pure photon or pure electron kernels among them and one can rewrite them easily through the 1- and 2-particle kernels. Then we use equations (\ref{eq:m_chaos}) and (\ref{eq:exch_f}) to complete the transformation.

\subsubsection{Closure of the Bogolyubov hierarchy}

Substituting equation (\ref{eq:ker2}) to equation (\ref{eq:ker1}), we get:
\beq
& \strut\disp \rho_{s_1}^{s'_1}\left({{\vk'_1}\atop {\vk_1}}\left|
\frac {T_0}{2}\right)-\rho_{s_1}^{s'_1}\left({{\vk'_1}\atop {\vk_1}}\right|-
\frac {T_0}{2}\right)=i
\frac{\alpha}{2\pi}\sum_{n,n'}\int
\frac {\d Y\d Z}{R}\frac {\d Y'\d Z'}{R'}\frac {\d\vk}{k}\frac {\d\vk'}{k'}
 & \nonumber \\
& \strut\disp 
\times\delta\left(n,Y,Z,\vk\;|\;n',Y',Z',\vk'\right)
M_{\sigma s}^{\sigma's'}\left({{n'Y'Z'}\atop {nYZ}}\left|{{\vk'}\atop {\vk}}
\right)  \right. & \nonumber \\
& \strut\disp 
\times\left[k'\delta_{s'_1}^{s'}(\vk'-\vk'_1)\rho_{s_1\sigma'n'}^
{s\sigma n}\left({{\vk}\atop {\vk_1}}\left|{{YZ}\atop {Y'Z'}}\right)-
k\delta_{s_1}^s(\vk-\vk_1)\rho_{s'\sigma'n'}^{s'_1\sigma n}\left({{\vk'_1}\atop
{\vk'}}\right|{{YZ}\atop {Y'Z'}}\right)\right]  & \nonumber \\
& \strut\disp -
\frac {\alpha^2}{(2\pi)^2}
\sum_{n,n',n'',n'''}\int\frac {\d Y\d Z}{R}\frac {\d Y'\d Z'}{R'}
\frac {\d Y''\d Z''}{R''}\frac {\d Y'''\d Z'''}{R'''}\frac {\d\vk}{k}
\frac {\d\vk'}{k'}\frac {\d\vk''}{k''}\frac {\d\vk'''}{k'''} 
 & \nonumber \\
& \strut\disp 
\times\delta\left(n,Y,Z,\vk\;|\;n',Y',Z',\vk'\right)
\delta\left(n'',Y'',Z'',\vk''\;|\;n''',Y''',Z''',\vk'''\right)
 &\nonumber \\
& \strut\disp 
\times M_{\sigma s}^{\sigma's'}\left({{n'Y'Z'}\atop {nYZ}}\left|
{{\vk'}\atop {\vk}}\right)M_{\sigma''s''}^{\sigma'''s'''}\left({{n'''Y'''Z'''}\atop
{n''Y''Z''}}\right|{{\vk'''}\atop {\vk''}}\right)  
& \nonumber \\
& \strut\disp 
\times\left\{
\delta_{s'_1}^{s'}(\vk'-\vk'_1)\left[k'k'''R'''\delta_{sn\sigma}^
{s'''n'''\sigma'''}(\vk-\vk''')\delta(\vpz-\vpz''')\rho_{s_1\sigma' n'}^
{s''\sigma''n''}\left({{\vk''}\atop {\vk_1}}\left|{{Y''Z''}\atop {Y'Z'}}\right)
\right.\right.\right. 
& \nonumber \\
& \strut\disp 
-k'k''R''\delta_{s_1n'\sigma'}^{s''n''\sigma''}(\vk_1-\vk'')
\delta(\vpz'-\vpz'')\rho_{s'''\sigma'''n'''}^{s\sigma n}\left({{\vk}\atop
{\vk'''}}\left|{{YZ}\atop {Y'''Z'''}}\right)  \right. & \nonumber \\
& \strut\disp +k'R'''\delta_{n\sigma}^{n'''\sigma'''}(\vpz'''-\vpz)
\rho_{s_1s'''\sigma' n'}^{s s''\sigma''n''}\left({{\vk\vk''}\atop
{\vk_1\vk'''}}\left|{{Y''Z''}\atop {Y'Z'}}\right)+k'k'''\delta_s^{s'''}
(\vk-\vk''')\rho_{s_1\sigma' n'\sigma'''n'''}^{s''\sigma n\sigma''n''}
\left({{\vk''}\atop {\vk_1}}\right|{{YZY''Z''}\atop {Y'Z'Y'''Z'''}}\right) 
 & \nonumber \\
& \strut\disp 
\left. -k'k''\delta_{s_1}^{s''}(\vk_1-\vk'')
\rho_{s'''\sigma'n'\sigma'''n'''}^{s\sigma n\sigma''n''}\left({{\vk}\atop
{\vk'''}}\left|{{YZY''Z''}\atop {Y'Z'Y'''Z'''}}\right)-k'R''
\delta_{n'\sigma'}^{n''\sigma''}(\vpz'-\vpz'')\rho_{s_1s'''\sigma n}^
{ss''\sigma'''n'''}\left({{\vk\vk''}\atop {\vk_1\vk'''}}\right|{{Y'''Z'''}\atop
{YZ}}\right)\right] 
& \nonumber \\
& \strut\disp 
+\delta_{s_1}^s(\vk-\vk_1)\left[-kk'''R'''\delta_{s_1'n\sigma}^
{s'''n'''\sigma'''}(\vk'''-\vk_1')\delta(\vpz-\vpz''')\rho_{s'\sigma' n'}^
{s''\sigma''n''}\left({{\vk''}\atop {\vk'}}\left|{{Y''Z''}\atop {Y'Z'}}\right) 
\right.\right.  
& \nonumber \\
& \strut\disp 
+kk''R''\delta_{s'}^{s''}(\vk'-\vk'')\delta_(n'\sigma')^
{n''\sigma''}(\vpz'-\vpz'')\rho_{s'''\sigma''' n'''}^{s_1'\sigma n}
\left({{\vk_1'}\atop {\vk'''}}\left|{{YZ}\atop {Y'''Z'''}}\right)  \right.
 & \nonumber \\
& \strut\disp 
-kR'''\delta_{n\sigma}^{n'''\sigma'''}(\vpz-\vpz''')
\rho_{s's'''\sigma'n'}^{s_1's''\sigma''n''}\left({{\vk_1'\vk''}\atop
{\vk'\vk'''}}\left|{{Y''Z''}\atop {Y'Z'}}\right)+kR''\delta_{n'\sigma'}^
{n''\sigma''}(\vpz'-\vpz'')\rho_{s's'''\sigma n}^{s_1's''\sigma''' n'''}
\left({{\vk_1'\vk''}\atop {\vk'\vk'''}}\right|{{Y'''Z'''}\atop {YZ}}\right) 
 & \nonumber \\
& \strut\disp 
\left.\left. -kk'''\delta_{s_1'}^{s'''}(\vk_1'-\vk''')
\rho_{s\sigma n}^{s'\sigma' n'}\left({{\vk''}\atop {\vk'}}\left|{{Y Z Y'' Z''}
\atop {Y'Z'Y'''Z'''}}\right)+kk''\delta_{s'}^{s''}(\vk'-\vk'')
\rho_{s'''\sigma'n'\sigma'''n'''}^{s_1'\sigma n\sigma''n''}\left({{\vk_1'}\atop
{\vk'''}}\right|{{YZY''Z''}\atop {Y'Z'Y'''Z'''}}\right)\right]\right\} , &
\eeq
where we introduced a symbol for the product of several Kronecker's deltas:
$\delta_{\alpha_1...\alpha_N}^{\beta_1...\beta_N}\equiv\prod_{i=1}^N
\delta_{\alpha_i}^{\beta_i},$
a symbol for the product of Kronecker's deltas and a $\delta$-function:
$\delta_{\alpha_1...\alpha_N}^{\beta_1...\beta_N}(a)\equiv\delta_{\alpha_1...
\alpha_N}^{\beta_1...\beta_N}\delta(a)$, 
and  $\vpz\equiv(0,Y,Z)$ is the electron momentum.

Using equations (\ref{eq:m_chaos}), (\ref{eq:exch_f}) and (\ref{eq:exch_e}), one can transform 2- and 3-particle kernels in the {rhs} of the equation to the 1-particle kernels. Then using equation (\ref{eq:nucl2df2}), the equation can be represented in terms of the coherency matrix. After some algebra we get:
\beq \label{eq:ku_1}
& \strut\disp \rho_{s_1}^{s'_1}\left({{\vk'_1}\atop {\vk_1}}\left|
\frac {T_0}{2}\right)-\rho_{s_1}^{s'_1}\left({{\vk'_1}\atop {\vk_1}}\right|-
\frac {T_0}{2}\right) & \nonumber \\
& \strut\disp  =i
\frac {\alpha}{2\pi}\sum_{n,n'}\int
\frac {\d Y\d Z}{R}\frac {\d Y'\d Z'}{R'}\frac {\d\vk}{k}\frac {\d\vk'}{k'} 
\delta\left(n,Y,Z,\vk\;|\;n',Y',Z',\vk'\right)
\cA_1 
& \nonumber \\
& \strut\disp 
+
\frac {\alpha^2}{4\pi^2}
\sum_{n,n',n'',n'''}\int\frac {\d Y\d Z}{R}\frac {\d Y'\d Z'}{R'}
\frac {\d Y''\d Z''}{R''}\frac {\d Y'''\d Z'''}{R'''}\frac {\d\vk}{k}
\frac {\d\vk'}{k'}\frac {\d\vk''}{k''}\frac {\d\vk'''}{k'''} 
& \nonumber \\
& \strut\disp 
\times\delta\left(n,Y,Z,\vk\;|\;n',Y',Z',\vk'\right)
M_{\sigma s}^{\sigma's'}\left({{n'Y'Z'}\atop {nYZ}}\left|{{\vk'}\atop {\vk}}\right)  \right. 
& \nonumber \\
& \strut\disp 
\times\delta\left(n'',Y'',Z'',\vk''\;|\;n''',Y''',Z''',\vk'''\right)
M_{\sigma''s''}^{\sigma'''s'''}\left({{n'''Y'''Z'''}\atop{n''Y''Z''}}\left|{{\vk'''}\atop {\vk''}}\right)  \right. 
& \nonumber \\
& \strut\disp \times\left\{
\cB_1+\cB_2+\cB_3+\cB_4 +\cB_5+\cB_6\right\}, &
\eeq
where
\beq
& \strut\disp \cA_1=M_{\sigma s}^{\sigma' s'}\left({{n'Y'Z'}\atop {nYZ}}\left|
{{\vk'}\atop {\vk}}\right)R'\delta_n^{n'}(\vpz-\vpz')\rho_{\sigma'n'}^{\sigma}
(\vpz)\left[k'\delta(\vk'-\vk'_1)\delta(\vk-\vk_1)\left(k_1
\delta_{s_1'}^{s_1}\rho_{s_1}^{s}(\vk_1)-k\delta_{s_1}^s\rho_{s'}^{s'_1}
(\vk')\right)\right], \right. & \\
& \strut\disp \cB_1=k'k'''\delta(\vk''-\vk''')\delta_n^{n'''}(\vpz-\vpz''')
\delta_{n'}^{n''}(\vpz'-\vpz'')\rho_{s'''}^{s''}(\vk''')\left(R'R'''
\delta_\sigma^{\sigma'''}\rho_{\sigma'n'}^{\sigma''}(\vpz')-RR''
\delta_{\sigma'}^{\sigma''}\rho_{\sigma n}^{\sigma'''}(\vpz)\right) 
 & \nonumber \\
& \strut\disp \times\left(k\delta(\vk_1'-\vk')\delta_{s_1}^s(\vk-\vk_1)
\rho_{s'}^{s'_1}(\vk')-k_1\delta(\vk_1-\vk)\delta_{s'_1}^{s'}(\vk'-\vk'_1)
\rho_{s_1}^{s}(\vk_{1})\right), & \\
& \strut\disp \cB_2=R'\delta_n^{n'}(\vpz-\vpz')\rho_{\sigma'n'}^{\sigma}(\vpz')
R'''\delta_{n'''}^{n''}(\vpz'''-\vpz'')\rho_{\sigma'''n'''}^{\sigma''}(\vpz''')
 & \nonumber \\
&\strut\disp \times\left[\delta_{s'_1}^{s'}(\vk'-\vk'_1)\left(k'k''k'''
\delta_{s_1}^{s''}(\vk_1-\vk'')\delta(\vk-\vk''')\rho_{s'''}^{s}(\vk''')-
k_1k'k'''\delta_s^{s'''}(\vk-\vk''')\delta(\vk''-\vk_1)\rho_{s_1}^{s''}(\vk_1)
\right) \right. & \nonumber \\
& \strut\disp \left. +\delta_{s_1}^s(\vk-\vk_1)\left(kk'k''\delta_{s'_1}^
{s'''}(\vk'_1-\vk''')\delta(\vk'-\vk'')\rho_{s'}^{s''}(\vk')-kk''k'''
\delta_{s'}^{s''}(\vk'-\vk'')\delta(\vk'_1-\vk''')\rho_{s'''}^{s'_{1}}(\vk''')
\right)\right], & \\
& \strut\disp \cB_3=\delta_{s_1}^s(\vk-\vk_1)kk'R'R'''\delta(\vk'-\vk'')
\delta(\vk'_1-\vk''')\delta_{n'}^{n''}(\vpz'-\vpz'')\delta_n^{n'''}
(\vpz-\vpz''')\rho_{s'}^{s''}(\vk')\rho_{\sigma'n'}^{\sigma''}(\vpz') 
 & \nonumber \\
& \strut\disp \times\left[k'''\delta_{\sigma}^{\sigma'''}\rho_{s'''}^{s'_1}
(\vk''')+k'''\delta_{s'_1}^{s'''}\left(\delta_\sigma^{\sigma'''}-
\rho_{\sigma'''n'''}^\sigma(\vpz''')\right)\right], & \\
& \strut\disp \cB_4=-k_1k'k'''R'R'''\delta_{s'_1}^{s'}(\vk'-\vk'_1)
\delta(\vk-\vk''')\delta(\vk_1-\vk'')\delta_n^{n'''}(\vpz-\vpz''')
\delta_{n'}^{n''}(\vpz'-\vpz'')\rho_{s_1}^{s''}(\vk_1)\rho_{\sigma'n'}^
{\sigma''}(\vpz')  & \nonumber \\
& \strut\disp \times\left[\delta_\sigma^{\sigma'''}\rho_{s'''}^s(\vk''')+
\delta_s^{s'''}\left(\delta_\sigma^{\sigma'''}-\rho_{\sigma'''n'''}^{\sigma}
(\vpz''')\right)\right], & \\
& \strut\disp \cB_5=k'k'''\delta(\vk-\vk''')\delta(\vk_1-\vk'')\delta_{s'_1}^
{s'}(\vk'-\vk'_1)\delta_{n}^{n'''}(\vpz-\vpz''')\delta_{n'}^{n''}(\vpz'-\vpz'')
\rho_{s'''}^s(\vk''')  & \nonumber \\
& \strut\disp \times\left[RR''k_{1}\delta_{\sigma'}^{\sigma''}\rho_{s_1}^{s''}
(\vk_1)\rho_{\sigma n}^{\sigma'''}(\vpz)+k''R'''\delta_{s_1}^{s''}
\rho_{\sigma'''n'''}^{\sigma}(\vpz''')\left(R''\delta_{\sigma'}^{\sigma''}-R'
\rho_{\sigma'n'}^{\sigma''}(\vpz')\right)\right], & \\
& \strut\disp \cB_6=-kk'''\delta(\vk_1'-\vk''')\delta(\vk'-\vk'')
\delta_{s_1}^{s}(\vk-\vk_1)\delta_n^{n'''}(\vpz-\vpz''')\delta_{n'}^{n''}
(\vpz'-\vpz'')\rho_{s'''}^{s'_1}(\vk''')  & \nonumber \\
& \strut\disp \times\left[RR''k'\delta_{\sigma'}^{\sigma''}\rho_{s'}^{s''}
(\vk')\rho_{\sigma n}^{\sigma'''}(\vpz)+k''R'''\delta_{s'}^{s''}
\rho_{\sigma'''n'''}^\sigma(\vpz''')\left(R''\delta_{\sigma'}^{\sigma''}-R'
\rho_{\sigma'n'}^{\sigma''}(\vpz')\right)\right]. &
\eeq

\subsubsection{Simplification of equation (\ref{eq:ku_1})}

The presence of the $\delta$-functions under the integrals in equation (\ref{eq:ku_1}) allows us to reduce a number of integrations. Let us define two singular measures $\mu_1$ and $\mu_2$:
\beq
& \strut\disp \d\mu_1=\frac {\d Y\d Z}{R}\frac {\d Y'\d Z'}{R'}\frac {\d\vk}{k}
\frac {\d\vk'}{k'}\delta\left(n,Y,Z,\vk\;|\;n',Y',Z',\vk'\right), & \\
& \strut\disp \d\mu_2=\frac {\d Y\d Z}{R}\frac {\d Y'\d Z'}{R'}
\frac {\d Y''\d Z''}{R''}\frac {\d Y'''\d Z'''}{R'''}\frac {\d\vk}{k}
\frac {\d\vk'}{k'}\frac {\d\vk''}{k''}\frac {\d\vk'''}{k'''}\times
& \nonumber \\
& \strut\disp
\times\delta\left(n,Y,Z,\vk\;|\;n',Y',Z',\vk'\right)
\delta\left(n'',Y'',Z'',\vk''\;|\;n''',Y''',Z''',\vk'''\right). &
\eeq
The terms from the {rhs} of equation (\ref{eq:ku_1}) can be written as:
\beq \label{eq:a1}
& \strut\disp \sum_{n,n'}\int\d\mu_1\cA_1=\sum_n\int\frac {\d Y\d Z}{R}
M_{\sigma s}^{\sigma' s'}\left({{nYZ}\atop {nYZ}}\left|{{\vk_1}\atop {\vk_1}}
\right)\rho_{\sigma'n}^\sigma(Z)\left[\delta_{s'_1}^{s'}\rho_{s_1}^{s}(\vk_1)-
\delta_{s_1}^{s}\rho_{s'}^{s'_1}(\vk_1)\right] \right. & \\ \label{eq:b1}
& \strut\disp \sum_{n,...,n'''}\int\d\mu_2\cB_1=\sum_{n,n'}\int
\frac {\d Y\d Z}{R}\frac {\d Y'\d Z'}{R'}\frac {\d\vk}{k}\delta(R'-R)M_{\sigma''s''}^
{\sigma''' s'''}\left({{nYZ}\atop {n'Y'Z'}}\left|{{\vk}\atop {\vk}}\right)
M_{\sigma s}^{\sigma' s'}\left({{n'Y'Z'}\atop {nYZ}}\right|{{\vk'_1}\atop {\vk_1}}
\right)\rho_{s'''}^{s''}(\vk)  & \nonumber \\
& \strut\disp \times\left[\delta_\sigma^{\sigma'''}\rho_{\sigma'n'}^{\sigma''}
(Z')-\delta_{\sigma'}^{\sigma''}\rho_{\sigma n}^{\sigma'''}(Z)\right]
\left[\delta_{s_1}^s\rho_{s'}^{s'_1}(\vk'_1)-\delta_{s'_1}^{s'}\rho_{s_1}^{s}
(\vk_1)\right], & 
\eeq
\beq \label{eq:b2}
& \strut\disp \sum_{n,...,n'''}\int\d\mu_2\cB_2=\sum_{n,n'}\int
\frac {\d Y\d Z}{R}\frac {\d Y'\d Z'}{R'}\frac {\d\vk}{k}\delta(\vk_1-\vk)
\rho_{\sigma'n}^\sigma(Z)\rho_{\sigma'''n'}^{\sigma''}(Z')\left[
M_{\sigma s}^{\sigma's'_1}\left({{nYZ}\atop {nYZ}}\left|{{\vk_1}\atop {\vk}}
\right)M_{\sigma''s''}^{\sigma'''s'''}\left({{n'Y'Z'}\atop {n'Y'Z'}}\right|{{\vk}
\atop {\vk_1}}\right)  \right. & \nonumber \\
& \strut\disp \left. \times\left(\delta_{s_1}^{s''}\rho_{s'''}^s(\vk)-
\delta_s^{s'''}\rho_{s_1}^{s''}(\vk_1)\right)+M_{\sigma s_1}^{\sigma's'}
\left({{nYZ}\atop {nYZ}}\left|{{\vk}\atop {\vk_1}}\right)M_{\sigma''s''}^
{\sigma'''s'''}\left({{n'Y'Z'}\atop {n'Y'Z'}}\right|{{\vk_1}\atop {\vk}}\right)
\left(\delta_{s'_1}^{s'''}\rho_{s'}^{s''}(\vk)-\delta_{s'}^{s''}\rho_{s'''}^
{s'_1}(\vk_1)\right)\right], &
\eeq
\beq\label{eq:b3}
& \strut\disp \sum_{n,...,n'''}\int\d\mu_2\cB_3=\sum_{n,n'}\int\frac {\d Y\d Z}{R}
\frac {\d Y'\d Z'}{R'}\frac {\d\vk}{k}\delta\left(n,Y,Z,\vk_1\;|\;n',Y',Z',\vk\right)
  & \nonumber \\
& \strut\disp \times M_{\sigma s_1}^{\sigma's'}\left({{n'Y'Z'}\atop {nYZ}}\left|
{{\vk}\atop {\vk_1}}\right)M_{\sigma''s''}^{\sigma''' s'''}\left({{nYZ}\atop
{n'Y'Z'}}\right|{{\vk_1}\atop {\vk}}\right)\rho_{s'}^{s''}(\vk)\rho_{\sigma'n'}^
{\sigma''}(Z')\left[\delta_{\sigma}^{\sigma'''}\rho_{s'''}^{s'_1}(\vk_1)+
\delta_{s'_1}^{s'''}\left(\delta_\sigma^{\sigma'''}-\rho_{\sigma'''n}^{\sigma}
(Z)\right)\right], & 
\eeq
\beq\label{eq:b4}
& \strut\disp \sum_{n,...,n'''}\int\d\mu_2\cB_4=-\sum_{n,n'}\int\frac {\d Y\d Z}{R}
\frac {\d Y'\d Z'}{R'}\frac {\d\vk}{k}\delta\left(n,Y,Z,\vk\;|\;n',Y',Z',\vk_{1}\right)  
& \nonumber \\
& \strut\disp 
\times M_{\sigma s}^{\sigma' s'_1}\left({{n'Y'Z'}\atop {nYZ}}\left|
{{\vk_1}\atop {\vk}}\right)M_{\sigma'' s''}^{\sigma'''s'''}\left({{nYZ}\atop
{n'Y'Z'}}\right|{{\vk}\atop {\vk_1}}\right)\rho_{s_1}^{s''}(\vk_1)
\rho_{\sigma'n'}^{\sigma''}(Z')\left[\delta_{\sigma}^{\sigma'''}\rho_{s'''}^s
(\vk)+\delta_s^{s'''}\left(\delta_\sigma^{\sigma'''}-\rho_{\sigma'''n}^{\sigma}
(Z)\right)\right], & 
\eeq 
\beq\label{eq:b5}
& \strut\disp \sum_{n,...,n'''}\int\d\mu_2\cB_5=\sum_{n,n'}\int\frac {\d Y\d Z}{R}
\frac {\d Y'\d Z'}{R'}\frac {\d\vk}{k}\delta\left(n,Y,Z,\vk\;|\;n',Y',Z',\vk_{1}\right)  
& \nonumber \\
& \strut\disp 
\times M_{\sigma s}^{\sigma's'_1}\left({{n'Y'Z'}\atop {nYZ}}\left|
{{\vk_1}\atop {\vk}}\right)M_{\sigma'' s''}^{\sigma''' s'''}\left({{nYZ}\atop
{n'Y'Z'}}\right|{{\vk}\atop {\vk_1}}\right)\rho_{s'''}^{s}(\vk)\left[
\delta_{\sigma'}^{\sigma''}\rho_{s_1}^{s''}(\vk_1)\rho_{\sigma n}^{\sigma'''}
(Z)+\delta_{s_1}^{s''}\rho_{\sigma'''n}^{\sigma}(Z)\left(\delta_{\sigma'}^
{\sigma''}-\rho_{\sigma'n'}^{\sigma''}(Z')\right)\right], & 
\eeq 
\beq\label{eq:b6}
& \strut\disp \sum_{n,...,n'''}\int\d\mu_2\cB_6=-\sum_{n,n'}\int\frac {\d Y\d Z}{R}
\frac {\d Y'\d Z'}{R'}\frac {\d\vk}{k}\delta\left(n,Y,Z,\vk\;|\;n',Y',Z',\vk_{1}\right)  
& \nonumber \\
& \strut\disp 
\times M_{\sigma s_1}^{\sigma' s'}\left({{nYZ}\atop {n'Y'Z'}}\left|
{{\vk}\atop {\vk_1}}\right)M_{\sigma''s''}^{\sigma''' s'''}\left({{n'Y'Z'}\atop
{nYZ}}\left|{{\vk_1}\atop {\vk}}\right)\rho_{s'''}^{s'_1}(\vk_1)\right[
\delta_{\sigma'}^{\sigma''}\rho_{s'}^{s''}(\vk)\rho_{\sigma n'}^{\sigma'''}(Z')
+\delta_{s'}^{s''}\rho_{\sigma'''n'}^{\sigma}(Z')\left(\delta_{\sigma'}^
{\sigma''}-\rho_{\sigma'n}^{\sigma''}(Z)\right)\right]. &
\eeq
We use the {rhs} of equations (\ref{eq:a1})--(\ref{eq:b6}) in the {rhs} of the final kinetic equation. 
One can notice that the  {rhs} of expression (\ref{eq:b1}) vanishes after summation over the electrons spin states and the photons polarizations.

\subsubsection{Transformation of the left-hand size of equation (\ref{eq:ku_1})}

It is necessary to rewrite the  {lhs} of equation (\ref{eq:ku_1}) in terms of the distribution function. Then one can rewrite  {lhs} through the differential operator. The later transformation is similar to the one, which is made in the quantum field theory for transition from the limited space-box to the infinite space. Finally we get:
\be
\rho_{s_1}^{s'_1}\left({{\vk'_1}\atop {\vk_1}}\left|
\frac {T_0}{2}\right)-\rho_{s_1}^{s'_1}\left({{\vk'_1}\atop {\vk_1}}\left|-
\frac {T_0}{2}\right)=T_0\DR {}{t}\rho_{s_1}^{s'_1}\left({{\vk'_1}\atop
{\vk_1}}\right|t\right)=T_0k_1\delta(\vk'_1-\vk_1)\DR {}{t}\rho_{s_1}^{s'_1}
(\vk_{1})=k_1\delta(\uk'_1-\uk_{1})\frac {2\pi}{c}\frac{\d r}{\d t}\rho_{s_1}^{s'_1}(\vk_1), \right.
\ee
where in the later transition the relation $\delta(k_{1}-k'_{1})=cT_0/(2\pi)$ is used. Then one can restore the dependence of the photon matrix on time and space coordinates. After rewriting the derivative over the line of sight as the full derivative, the  
{lhs} of the equation takes the covariant form
\be
k_1\delta(\uk'_1-\uk_{1})\frac {2\pi}{c}\frac{\d r}{\d t}\rho_{s_1}^{s'_1}(\vk_1)
\longmapsto
2\pi\delta(\uk_{1}-\uk'_{1})
\left(\frac{k_{1}}{c}\frac{\partial}{\partial t}+\vk_{1}\right)\rho_{s_1}^{s'_1}(\vk_1,\vr_1,t)\equiv
2\pi\delta(\uk_{1}-\uk'_{1})\uk_{1}\underline{\nabla}\rho_{s_1}^{s'_1}(\vk_1,\vr_1,t).
\ee

\section{Different forms of the kinetic equation}

\subsection{Kinetic equation for coherency matrix}

\subsubsection{The general form of the kinetic equation}

Now one can write the final form of the kinetic equation. In the most general case, we formulate it for the coherency matrix, where the polarization of  electrons is taken into account (i.e. for the situation when there can be non-trivial spin-distribution of the electron gas).
The equation for the case of polarized electrons is:
\beq \label{eq:ku_e_pol}
& \strut\disp \uk_1\unb\rho_{s_1}^{s'_1}(\vk_1,\vr_1,t)=I_{1}+I_{2}+I_{3},&
\eeq
where
\beq 
&I_{1}=i\alpha \strut\disp 
\frac {1}{(2\pi)^2}\sum_n\int\frac {\d Y\d Z}{R}\rho_{\sigma'n}^\sigma(Z)
\left[\rho_{s_1}^s(\vk_1)M_{\sigma s}^{\sigma's'_1}
\left({{nYZ}\atop {nYZ}}\left|{{\vk_1}\atop {\vk_1}}\right)-
\rho_{s'}^{s'_1}(\vk_1)M_{\sigma s_1}^{\sigma's'}\left({{nYZ}\atop {nYZ}}\right|
{{\vk_1}\atop {\vk_1}}\right)\right],&
\eeq
\beq& \strut\disp 
I_{2}=\frac {\alpha^2}{(2\pi)^3}\sum_{n,n'}\int
\frac {\d Y\d Z}{R}\frac {\d Y'\d Z'}{R'}\frac {\d\vk}{k}\delta(k-k_1)
\delta(k\cos\theta-k_1\cos\theta_1)\rho_{\sigma'n}^{\sigma}(Z)
\rho_{\sigma'''n'}^{\sigma''}(Z')  
& \nonumber \\
& \strut\disp 
\times\left[2M_{\sigma s}^{\sigma's'_{1}}\left({{nYZ}\atop {nYZ}}
\left|{{\vk_1}\atop {\vk}}\right)M_{\sigma''s_1}^{\sigma'''s'''}\left({{n'Y'Z'}
\atop {n'Y'Z'}}\right|{{\vk}\atop {\vk_1}}\right)\rho_{s'''}^s(\vk)-
M_{\sigma s}^{\sigma's'_1}\left({{nYZ}\atop {nYZ}}\left|{{\vk_1}\atop {\vk}}
\right)M_{\sigma''s''}^{\sigma'''s}\left({{n'Y'Z'}\atop {n'Y'Z'}}\right|{{\vk}\atop
{\vk_1}}\right)\rho_{s_1}^{s''}(\vk_1) \right. 
& \nonumber \\
& \strut\disp 
\left. -M_{\sigma''s'}^{\sigma'''s'''}\left({{n'Y'Z'}\atop {n'Y'Z'}}
\left|{{\vk_1}\atop {\vk}}\right)M_{\sigma s_1}^{\sigma's'}\left({{nYZ}\atop
{nYZ}}\right|{{\vk}\atop {\vk_1}}\right)\rho_{s'''}^{s'_1}(\vk_1)\right],
&\eeq
\beq& \strut\disp 
I_{3}= 
\frac{\alpha^2}{(2\pi)^3}\sum_{n,n'}\int
\frac{\d Y\d Z}{R}\frac {\d Y'\d Z'}{R'}\frac {\d\vk}{k}\delta(R+k-R'-k_1)
\delta(Z+k\cos\theta-Z'-k_1\cos\theta_1)  
& \nonumber \\
& \strut\disp 
\times\left\{
\left[\delta_\sigma^{\sigma'''}\rho_{\sigma'n}^{\sigma''}(Z)-
\delta_{\sigma'}^{\sigma''}\rho_{\sigma n'}^{\sigma'''}(Z')\right] 
\left[M_{\sigma s_1}^{\sigma's'}\left({{nYZ}\atop {n'Y'Z'}}
\left|{{\vk}\atop {\vk_1}}\right)M_{\sigma'' s''}^{\sigma''' s'''}
\left({{n'Y'Z'}\atop {nYZ}}\right|{{\vk_1}\atop {\vk}}\right)\rho_{s'''}^{s'_1}
(\vk_1) \right. \right. 
&\\
& \strut\disp 
\left.+M_{\sigma' s''}^{\sigma s'_1}\left({{n'Y'Z'}\atop {nYZ}}\left|{{\vk_1}
\atop {\vk}}\right)M_{\sigma'''s'''}^{\sigma''s'}\left({{nYZ}\atop {n'Y'Z'}}
\right|{{\vk}\atop {\vk_1}}\right)\rho_{s_1}^{s'''}(\vk_1)\right]
\rho_{s'}^{s''}(\vk) 
& \nonumber \\
& \strut\disp 
+2M_{\sigma s_1}^{\sigma's'}\left({{nYZ}\atop {n'Y'Z'}}\left|{{\vk}
\atop {\vk_1}}\right)M_{\sigma''s''}^{\sigma''' s'_1}\left({{n'Y'Z'}\atop {nYZ}}
\right|{{\vk_1}\atop {\vk}}\right)\rho_{s'}^{s''}(\vk)
\rho_{\sigma'n}^{\sigma''}(Z)\left[\delta_\sigma^{\sigma'''}-
\rho_{\sigma'''n'}^\sigma(Z')\right] 
& \nonumber \\
& \strut\disp 
-\rho_{\sigma'''n'}^\sigma(Z')\left[\delta_{\sigma'}^{\sigma''}-
\rho_{\sigma'n}^{\sigma''}(Z)\right]  
& \nonumber \\
& \strut\disp 
\left.\times\left[M_{\sigma s_1}^{\sigma' s''}\left({{nYZ}\atop
{n'Y'Z'}}\left|{{\vk}\atop {\vk_1}}\right)M_{\sigma'' s''}^{\sigma''' s'''}
\left({{n'Y'Z'}\atop {nYZ}}\right|{{\vk_1}\atop {\vk}}\right)\rho_{s'''}^{s'_1}
(\vk_1)+M_{\sigma's}^{\sigma'''s'_1}\left({{n'Y'Z'}\atop {nYZ}}\left|{{\vk_1}\atop
{\vk}}\right)M_{\sigma s''}^{\sigma''s}\left({{nYZ}\atop {n'Y'Z'}}\right|{{\vk}
\atop {\vk_1}}\right)\rho_{s_1}^{s''}(\vk_1)\right]\right\}.
&\nonumber\eeq
There are three terms in the {rhs} of equations (\ref{eq:ku_e_pol}). 
The first and the second terms describes redistribution only over polarization.
The last one describes the general redistribution of the photons over quantum states (energies, momentum directions and polarization).

\subsubsection{Equation for the case of non-polarized electrons}

The kinetic equation in the case non-polarized electrons one  can deduced from equation (\ref{eq:ku_e_pol}) by averaging over the spin states of the electrons. There is a relation between the distribution function of the electrons $f_{n}(Z)$ and the diagonal elements of the electron coherency matrix:
\be 
f_{n}(Z)=\rho_{1n}^{1}(Z)+\rho_{2n}^{2}(Z),
\ee
where
$\rho_{1n}^{1}(Z)=\rho_{2n}^{2}(Z).$ The distribution function is normalized to the total number of the electrons: 
\be 
\sum_{n}\int\d Z f_n(Z)=N_{{\rm e}}.
\ee
Then equation for the case of non-polarized electrons is:
\beq \label{eq:ku_e_nopol}
& \strut\disp 
\uk_1\unb\rho_{s_1}^{s'_1}(\vk_1,\vr_1,t)
=
J_{1}+J_{2}+J_{3}, &
\eeq
where
\beq& \strut\disp 
J_{1}=
i\frac{\alpha}
{(2\pi)^2}\sum_n\int\frac {\d Y\d Z}{R}\frac{f_n(Z)}{2}\left[
\rho_{s_1}^{s}(\vk_1)M_{\sigma s}^{\sigma s'_{1}}\left({{nYZ}\atop {nYZ}}\left|
{{\vk_1}\atop {\vk_1}}\right)-\rho_{s'}^{s'_1}(\vk_1)M_{\sigma
s_1}^{\sigma s'}\left({{nYZ}\atop {nYZ}}\right|{{\vk_1}\atop {\vk_1}}\right)\right],
&\eeq
\beq& \strut\disp 
J_{2}=
\frac {\alpha^2}{(2\pi)^3}\sum_{n,n'}\int
\frac {\d Y\d Z}{R}\frac {\d Y'\d Z'}{R'}\frac {\d\vk}{k}\delta(k-k_1)
\delta(k\cos\theta-k_1\cos\theta_1)\frac{f_n(Z)}{2}\frac{f_{n'}(Z')}{2}
  & \nonumber \\
& \strut\disp \times\left[2M_{\sigma s}^{\sigma s'_{1}}\left({{nYZ}\atop {nYZ}}
\left|{{\vk_1}\atop {\vk}}\right)M_{\sigma''s_1}^{\sigma''s'''}\left({{n'Y'Z'}
\atop {n'Y'Z'}}\right|{{\vk}\atop {\vk_1}}\right)\rho_{s'''}^s(\vk)-
M_{\sigma s}^{\sigma s'_1}\left({{nYZ}\atop {nYZ}}\left|{{\vk_1}\atop {\vk}}
\right)M_{\sigma''s''}^{\sigma''s}\left({{n'Y'Z'}\atop {n'Y'Z'}}\right|{{\vk}\atop
{\vk_1}}\right)\rho_{s_1}^{s''}(\vk_1)  \right. & \nonumber \\
& \strut\disp \left. -M_{\sigma''s'}^{\sigma''s'''}\left({{n'Y'Z'}\atop {n'Y'Z'}}
\left|{{\vk_1}\atop {\vk}}\right)M_{\sigma s_1}^{\sigma s'}\left({{nYZ}\atop
{nYZ}}\right|{{\vk}\atop {\vk_1}}\right)\rho_{s'''}^{s'_1}(\vk_1)\right],
&\eeq
\beq& \strut\disp 
J_{3}=
\frac {\alpha^2}{(2\pi)^3}\sum_{n,n'}\int
\frac {\d Y\d Z}{R}\frac {\d Y'\d Z'}{R'}\frac {\d\vk}{k}\delta(R+k-R'-k_1)
\delta(Z+k\cos\theta-Z'-k_1\cos\theta_1)  & \nonumber \\
& \strut\disp \times
\frac{1}{2}\left\{  \left[f_n(Z)-f_{n'}(Z') \right]
\left[T_{s''s_1}^{s'''s'}\rho_{s'''}^{s'_1}(\vk_1)+T_{s''s'''}^{s'_1s'}
\rho_{s_1}^{s'''}(\vk_{1})\right]\rho_{s'}^{s''}(\vk) 
+2T_{s''s_1}^{s'_{1}s'}\rho_{s'}^{s''}(\vk)f_n(Z)\left[1-\frac{f_n'(Z')}{2} \right]  \right. 
& \nonumber \\
& \strut\disp \left. -f_{n'}(Z')\left[1-\frac{f_n(Z)}{2}\right]\left[T_{s''s_1}^{s'''s''}
\rho_{s'''}^{s'_1}(\vk_1)+T_{ss''}^{s'_1s}\rho_{s_1}^{s''}(\vk_1)\right]
\right\} ,
&\eeq
where we use a notation for the product of pairs of the scattering matrix elements:
\be\label{T-matrix_def}
T_{jm}^{ik}\equiv M_{\sigma'j}^{\sigma i}\left({{n'Y'Z'}\atop {nYZ}}\left|
{{\vk_1}\atop {\vk}}\right)M_{\sigma m}^{\sigma'k}\left({{nYZ}\atop {n'Y'Z'}}
\right|{{\vk}\atop {\vk_1}}\right).
\ee
After averaging over the spin states of the electrons the last term in the {rhs} of the kinetic equation  simplifies. 
The $\delta$-functions under the integrals in the  {rhs} of equations (\ref{eq:ku_e_pol}) and (\ref{eq:ku_e_nopol}) 
can be use to reduce the number of integrations and to simplify the sum.

\subsubsection{Equation for the case of non-polarized rarefied electron gas}

Another form of the kinetic equations can be obtained in the case of rarefied electron gas. 
Neglecting in equation (\ref{eq:ku_e_nopol}) the terms containing squares of the electron distribution function, we gets  
\beq \label{eq:ku_e_nopol_nodeg}
& \strut\disp \uk_1\unb\rho_{s_1}^{s'_1}(\vk_1,\vr_1,t)=
K_{1}+K_{2}+K_{3}, &
\eeq
where
\beq&\strut\disp 
K_{1}=i\frac {\alpha}
{(2\pi)^2}\sum_n\int\frac {\d Y\d Z}{R}\frac{f_n(Z)}{2}\left[
\rho_{s_1}^{s}(\vk_1)M_{\sigma s}^{\sigma s'_{1}}\left({{nYZ}\atop {nYZ}}\left|
{{\vk_1}\atop {\vk_1}}\right)-
\rho_{s'}^{s'_1}(\vk_1)M_{\sigma
s_1}^{\sigma s'}\left({{nYZ}\atop {nYZ}}\right|{{\vk_1}\atop {\vk_1}}\right)\right],
&\eeq
\beq&
K_{2}=0,
&\eeq
\beq& \strut\disp 
K_{3}=\frac {\alpha^2}{(2\pi)^3}\sum_{n,n'}\int
\frac {\d Y\d Z}{R}\frac {\d Y'\d Z'}{R'}\frac {\d\vk}{k}\delta(R+k-R'-k_1)
\delta(Z+k\cos\theta-Z'-k_1\cos\theta_1)\times & \nonumber \\
& \strut\disp \times\frac{1}{2}\left\{
f_{n}(Z)\rho_{s'}^{s''}(\vk)\left[ 2T_{s''s_{1}}^{s'_{1}s'}+T_{s''s_{1}}^{s'''s'}\rho_{s'''}^{s'_{1}}(\vk_{1})
+T_{s''s'''}^{s'_{1}s'}\rho_{s_{1}}^{s'''}(\vk_{1})\right]-\right. 
& \nonumber \\
& \strut\disp
\left. -f_{n'}(Z')\left[ \rho_{s'''}^{s'_{1}}(\vk_{1})\left(   T_{s''s_{1}}^{s'''s''}+T_{s''s_{1}}^{s'''s'}\rho_{s'}^{s''}(\vk)\right)+
\rho_{s_{1}}^{s'''}(\vk_{1})\left(T{^{s'_{1}}_{s}}{^{s}_{s'''}}+T_{s''s'''}^{s'_{1}s'}\rho_{s'}^{s''}(\vk)\right)
\right]
\right\}. &
\eeq
We notice that the first term of equation (\ref{eq:ku_e_nopol}) has not changed, while the second term has now disappeared.

\subsubsection{Kinetic equation  in terms of two polarization modes}

In the case of non-polarized rarefied electron gas, equation (\ref{eq:ku_e_nopol_nodeg}) can be simplified further 
if one assumes the absence of correlations between the two linear polarization modes:
$\rho_{1}^{2}(\vk)=\rho_{2}^{1}(\vk)=0.$
Then one can use only one polarization index for the diagonal elements of the coherency matrix:
$\rho_{i}(\vk)\equiv\rho_{i}^{i}(\vk).$
The kinetic equation then get the form:
\beq \label{eq:ku_e_nopol_nodeg_2m}
& \strut\disp \uk_1\unb\rho_{s_1}(\vk_1,\vr_1,t)=
L_{1}+L_{2}+L_{3},
&\eeq
where $L_{1}=0$ and $L_{2}=0$, and 
\beq&  \strut\disp
L_{3}=\frac {\alpha^2}{(2\pi)^3}\sum_{n,n'}\int
\frac {\d Y\d Z}{R}\frac {\d Y'\d Z'}{R'}\frac {\d\vk}{k}\delta(R+k-R'-k_1)
\delta(Z+k\cos\theta-Z'-k_1\cos\theta_1)  & \nonumber \\
& \strut\disp \times
T{^{s_1}_{s}}{^{s}_{s_1}}\left\lbrace f_{n}(Z)\rho_{s}(\vk)\left[ 1+\rho_{s_{1}}(\vk_{1})\right]-
f_{n'}(Z')\rho_{s_{1}}(\vk_{1})\left[ 1+\rho_{s}(\vk)\right]  
\right\rbrace.
&\eeq
This form of the equation is obvious and can be written immediately using physical arguments \cite{PSM1989}.
In this case the two modes are considered independently and a possibility of correlation between their phases is not taken into account.

\subsection{Kinetic equation in terms of Stokes parameters}

Equations (\ref{eq:ku_e_pol}) and (\ref{eq:ku_e_nopol}) can be rewritten in  terms of Stokes parameters. Transformation to this form can be done using trivial linear transformation. Elements of the coherency matrix $\{\rho_j^i(\vk)\}$ and the Stokes vector $\vN=(n_{\rm I},n_{\rm Q},n_{\rm U},n_{\rm V})^{T}$ are connected by relations:  
\beq
& \strut\disp n_{\rm I}=(\rho_1^1+\rho_2^2)/2,\quad n_{\rm Q}=(\rho_1^1-\rho_2^2)/2,\quad
n_{\rm U}=(\rho_1^2+\rho_2^1)/2,\quad n_{\rm V}=i(\rho_1^2-\rho_2^1)/2, & \\
& \strut\disp \rho_1^1=n_{\rm I}+n_{\rm Q},\quad \rho_2^2=n_{\rm I}-n_{\rm Q},\quad
\rho_1^2=n_{\rm U}-in_{\rm V},\quad\rho_2^1=n_{\rm U}+in_{\rm V}. &
\eeq

\subsubsection{General equation for the case of polarized electrons}
\label{sec:genpol1}

Using equation (\ref{eq:ku_e_pol}), one can find the kinetic equation in the case of polarized electrons in terms of Stokes parameters:
\beq \label{eq:Sp_p}
& \strut\disp \uk_1\unb\vN_1=
I^{P}_{1}+I^{P}_{2}+I^{P}_{3}, &
\eeq
where
\beq \label{eq:Sp_p_I1}
&  \strut\disp 
I^{P}_{1}=
i\frac {\alpha}{(2\pi)^2}\sum_n\int
\frac {\d Y\d Z}{R}f^\sigma_{\sigma'n}(Z)\hat{\cF}\vN_1,
&\eeq
\beq&  \strut\disp 
I^{P}_{2}=
\frac {1}{2}\frac {\alpha^2}{(2\pi)^3}\sum_{n,n'}\int
\frac {\d Y\d Z}{R}\frac {\d Y'\d Z'}{R'}\frac {\d\vk}{k}\delta(k-k_1)
\delta(k\cos\theta-k_1\cos\theta_1)f^\sigma_{\sigma'n}(Z)
f^{\sigma''}_{\sigma'''n'}(Z')\left[\hat {\cF}'\vN+\hat{\cF}''\vN_1\right],
&\eeq
\beq  \label{eq:Sp_p_I3}
&  \strut\disp 
I^{P}_{3}=
\frac {1}{2}\frac {\alpha^2}{(2\pi)^3}\sum_{n,n'}\int
\frac {\d Y\d Z}{R}\frac {\d Y'\d Z'}{R'}\frac {\d\vk}{k}\delta(R+k-R'-k_1)
\delta(Z+k\cos\theta-Z'-k_1\cos\theta_1)  &  \\
& \strut\disp \times\left\{\left[\delta^{\sigma'''}_{\sigma}f^{\sigma''}_{\sigma'n}
(Z)-\delta^{\sigma''}_{\sigma'}f^{\sigma'''}_{\sigma n'}(Z') \right]
\hat{\cR}\vN_{1}+2f^{\sigma''}_{\sigma'n}(Z)\left[\delta^{\sigma'''}_{\sigma}-
f^\sigma_{\sigma'''n'}(Z') \right] \hat {\cR}'\vN-2f^\sigma_{\sigma'''n'}(Z')
\left[\delta^{\sigma''}_{\sigma'}-f^{\sigma''}_{\sigma'n}(Z) \right] \hat{\cR}''\vN_1
\right\}, \nonumber
&\eeq
where $\vN=\vN(\vk)$ and $\vN_{1}=\vN(\vk_{1})$ are the Stokes vectors, $\hat{\cF}$, $\hat{\cF}'$, $\hat{\cF}''$, $\hat{\cR}$, $\hat{\cR}'$ and $\hat{\cR}''$ are $4\times4$ complex matrices acting like linear operators from the real 4-dimensional space to the real 4-dimensional space.
The expressions for these matrices can be found in Appendix \ref{app_st1}.

\subsubsection{Equation for the case of non-polarized electrons}
\label{sec:nonpol2}

The equation for the case of non-polarized electrons can be derived from equation (\ref{eq:ku_e_nopol}):
\beq \label{eq:Sp_np}
& \strut\disp \uk_1\underline{\nabla}\vN_1=
J^{P}_{1}+J^{P}_{2}+J^{P}_{3}, &
\eeq
where
\beq \label{eq:Sp_np_J1} 
&  \strut\disp 
J^{P}_{1}=i\frac {\alpha}{(2\pi)^2}
\sum_n\int\frac {\d Y\d Z}{R}\frac{f_n(Z)}{2}\hat {\cF}\vN_1,
&\eeq
\beq&  \strut\disp 
J^{P}_{2}=\frac {1}{2}\frac {\alpha^2}{(2\pi)^3}\sum_{n,n'}\int\frac {\d Y\d Z}{R}
\frac {\d Y'\d Z'}{R'}\frac {\d\vk}{k}\delta(k-k_1)\delta(k\cos\theta-k_1
\cos\theta_1)\frac{f_{n}(Z)}{2}\frac{f_{n'}(Z')}{2}\left[\hat {\cF}'\vN+\hat{\cF}''\vN_1\right],
&\eeq
\beq \label{eq:Sp_np_J3}
&  \strut\disp 
J^{P}_{3}=\frac{1}{2}\frac{\alpha^2}{(2\pi)^3}\sum_{n,n'}\int
\frac {\d Y\d Z}{R}\frac {\d Y'\d Z'}{R'}\frac {\d\vk}{k}\delta(R+k-R'-k_{1})
\delta(Z+k\cos\theta-Z'-k_1\cos\theta_1)  & \nonumber \\
& \strut\disp \times\left\{\left[f_n(Z)-f_{n'}(Z')\right]\hat {\cR}\vN_1+f_n(Z)\left[1-\frac{f_{n'}(Z')}{2} \right]
\hat {\cR}'\vN-f_{n'}(Z')\left[1-\frac{f_n(Z)}{2} \right] \hat {\cR}''\vN_1\right\},
&\eeq
where $\hat{\cF}$, $\hat{\cF}'$, $\hat{\cF}''$, $\hat{\cR}$, $\hat{\cR}'$ and $\hat{\cR}''$ are $4\times4$ complex matrices 
that can be found in Appendix \ref{app_st2}. 

\subsubsection{Equation for the case of non-polarized rarefied electron gas}

We derive the equation for the case of rarefied electron gas by neglecting terms containing squared electron distribution functions in equation (\ref{eq:Sp_np}). The equation takes the following form:
\beq \label{eq:Sp_npr}
& \strut\disp \uk_1\underline{\nabla}\vN_1=
K^{P}_{1}+K^{P}_{2}+K^{P}_{3}, &
\eeq
where 
\beq& \strut\disp 
K^{P}_{1}=i\frac {\alpha}{(2\pi)^2}
\sum_n\int\frac {\d Y\d Z}{R}\frac{f_n(Z)}{2}\hat {\cF}\vN_1,
&\eeq
\beq&
K^{P}_{2}=0,
&\eeq
\beq& \strut\disp 
K^{P}_{3}=\frac{1}{2}\frac {\alpha^2}{(2\pi)^3}\sum_{n,n'}\int
\frac {\d Y\d Z}{R}\frac {\d Y'\d Z'}{R'}\frac {\d\vk}{k}\delta(R+k-R'-k_{1})
\delta(Z+k\cos\theta-Z'-k_1\cos\theta_1)\times & \nonumber \\
& \strut\disp \times\left[\left((f_{n}(Z)-f_{n'}(Z'))\hat{\cR}-f_{n'}(Z')\hat{\cR}''\right)\vN_{1}+f_{n}(Z)\hat{\cR}'\vN
\right],
&\eeq
where $\hat{\cF}$, $\hat{\cR}$, $\hat{\cR}'$ and $\hat{\cR}''$ are the same $4\times4$ complex matrices as in equations (\ref{eq:Sp_np})--(\ref{eq:Sp_np_J3}) and presented in Appendix \ref{app_st2}.

\section{Summary}

We have deduced a kinetic equation for Compton scattering of polarized radiation in magnetic field. Polarizations of photons and spin states of electrons, the induced scattering and the Pauli exclusion principle were taken into account. The equations are written for both the coherency matrix and  the Stokes parameters. Additional forms of the equations valid for the two polarization mode description of radiation is also derived. 
The equations for both polarized and non-polarized electrons were obtained. 
There are no significant (for the conditions in neutron stars atmospheres) limitations on the energies and the concentrations of the electrons and the photons.  The assumptions made are usual for the kinetic theory and related to the typical time scales of the problem and do not limit 
significantly the applicability range. 
 
The equations describe the interaction of radiation and electrons in strong magnetic field up to about $10^{16}$ G. 
There is no low limit on the B-field strength. 
At the same time, the derived equations become rather cumbersome in the case of weak magnetic field, because the electrons can occupy high Landau levels and there will be many terms in the {rhs} of the equation, where the summation over the Landau levels is carried out. 
On that other hand, in the case of strong magnetic field in neutron star atmospheres electrons typically occupy only ground Landau level 
(or only a few low levels), because of a rather low electron temperature and absence of high-energy photons. 
Therefore, the sums over $n$ and $n'$ has only a few terms, which simplifies the equations significantly. 

The most general form of the kinetic equation (\ref{eq:ku_e_pol}) has three terms. The second and the third terms there contain 
the products of two elements of the scattering matrix. These terms can be rewritten through the interaction cross sections. On the contrary, the first term in the {rhs} of equation (\ref{eq:ku_e_pol}) contains single elements of the scattering matrix. 
This term describes changes of the photon polarization with no corresponding changes in energy and the momentum direction. 
The polarization change term has the following form: the changes of the diagonal elements of the coherency matrix depend only on the non-diagonal elements and changes of the non-diagonal elements depend only on the diagonal ones. 
This term  describes the rotation of the polarization plane when radiation can be well described only by the Stokes parameters, and it disappears, if the kinetic equation is reformulated in terms of two polarization modes. It is possible that this term provides correction to the depolarization in the region of vacuum resonance, which most likely will not be large because of a small optical depth of this region.

The second term in the {rhs} of equation (\ref{eq:ku_e_pol}) describes redistribution of photons with only changes in polarization. 
It contains the products of the electron distribution functions and  disappears in the equation (\ref{eq:ku_e_nopol_nodeg}) for rarefied electron gas.
The last term in the {rhs} of equation (\ref{eq:ku_e_pol}) describes the general redistribution of photons over energy,  directions and polarizations. This term can be simplified significantly for the cases of non-polarized electrons (\ref{eq:ku_e_nopol}),  rarefied electron gas (\ref{eq:ku_e_nopol_nodeg}) and for the two-polarization mode description of radiation (\ref{eq:ku_e_nopol_nodeg_2m}). 
In the latter case this term is the only term in the {rhs} of the kinetic equation and coincides with previously known expressions. 
The derived equations form the basis for the construction of models of the radiative transfer in 
strongly magnetized neutron stars atmospheres and magnetospheres.

\acknowledgments
 
This study was supported by the CIMO grant TM-10-7326 (A.M.)  and the Academy of
Finland grant 127512  (J.P.). 
We are grateful to  Dmitry Yakovlev, Yuri Shibanov, Dmitry Rumyantsev, Valery Suleimanov, 
and an anonymous referee for a number of useful  comments.


\appendix

\section{Transformation from density matrix kernel to distribution function for the electron in magnetic field}
\label{Ritus}

The transformation from the kernel of density matrix to the distribution function for the case of charged particles in magnetic field is slightly more complicated than in the field-free case. We use the Landau gauge, with the electron momentum  is described by two continuous components $Y$ and $Z$ (corresponding to the $y$- and $z$-coordinates) and one discrete component, which corresponds to the $x$-coordinate. 
A transformation from the coordinate representation to the momentum representation cannot be done using Fourier transforms. Instead one should use a transformation based on the Ritus eigenfunctions which diagonalize the mass operator of electrons in presence of the external magnetic field \cite{Ritus1978}.

The kernel of the electron density matrix in the coordinate representation can be obtained from the kernel in the momentum representation:
\be \label{eq:rhofullvrvr}
\rho_\sigma^{\sigma'}(\vr\,',\vr)=\sum_{n,n'}\int\frac {\d Y\d Z}{\sqrt {R_n(Z)}}
\frac {\d Y'\d Z'}{\sqrt {R_{n'}(Z')}}e^{i(Y'y'-Yy+Z'z'-Zz)}\Phi_{n'}(u')\Phi_n(u)
\rho_{\sigma n}^{\sigma' n'}\left({Y'Z'}\atop {YZ}\right),
\ee
where $u=x-Y/b$, $u'=x'-Y'/b$ and $\Phi_n(x)$ are the functions related to the parabolic cylinder functions:
$$\Phi_n(x)=\frac {1}{\sqrt [4]{2\pi}\sqrt {n!}}D_n(x).$$
These functions compose the complete system satisfying the relations
$$\intl_{-\infty}^\infty\Phi_n(x)\Phi_{n'}(x)\d x=\delta_{n}^{n'}, \quad 
\sum_{n=0}^\infty\Phi_n(x)\Phi_n(x')=\delta(x'-x).$$

The inverse transformation from the kernel in coordinate representation to the kernel in momentum representation can be written in the 
following way:
\be \label{eq:rhoimpvr}
\rho_{\sigma n}^{\sigma' n'}\left({Y'Z'}\atop {YZ}\right)=
\frac {\sqrt {R_{n'}(Z')}\sqrt {R_n(Z)}}{(2\pi)^4}\int\d^3r\d^3r'
e^{-i(Y'y'-Yy+Z'z'-Zz)}\Phi_{n'}(u')\Phi_n(u)\rho_\sigma^{\sigma'}(\vr\,',\vr).
\ee

One can define the Wigner function for the electrons in magnetic field:
\beq
\rho_{\sigma n}^{\sigma'}(\vr,Y,Z)=\int\d v_y\d v_z
e^{i(Yv_y+Zv_z)}\int\d x'\Phi_n(u')
\rho_\sigma^{\sigma'}(x',y-v_y/2,z-v_z/2,x,y+v_y/2,z+v_z/2), 
\eeq
and rewrite it using the density matrix in momentum representation
\beq \label{eq:fundistrres}
& \strut\disp \rho_{\sigma n}^{\sigma'}(\vr,Y,Z)
=(2\pi)^2\sum_{n'}\!\!\int\!\!\frac {\d Y_1\d Z_1}{\sqrt {R_{n'}(Z_1)}}
\frac {\d Y_1'\d Z_1'}{\sqrt {R_{n}(Z'_1)}}
e^{i[(Y_1'-Y_1)y+(Z_1'-Z_1)z]}\delta\!\left(\!Y\!-\!\frac {Y_1\!+\!Y_1'}{2}\right)
& \nonumber \\
& \strut\disp
\times\delta\!\left(\!Z\!-\!\frac {Z_1\!+\!Z_1'}{2}\right)\Phi_{n'}(u)
\rho_{\sigma n}^{\sigma' n'}\left({Y_1'Z_1'}\atop {Y_1Z_1}\right). &
\eeq
The inverse transformation from Wigner function to the kernel in momentum representation is
\be \label{eq:revvrvr}
\rho_\sigma^{\sigma'}(\vr\,',\vr)=\sum_{n}\Phi_n(u')\int\d Y\d Z
e^{-i[(y-y')Y+(z-z')Z]}\rho_{\sigma n}^{\sigma'}(\vr,Y,Z).
\ee
From (\ref{eq:rhoimpvr}) and (\ref{eq:revvrvr}) one can obtain relation for kernel in momentum representation through the Wigner function:
\beq \label{eq:res-01}
& \strut\disp 
\rho_{\sigma n}^{\sigma' n'}\!\!\left({Y'Z'}\atop {YZ}\right)=
\frac {\sqrt {R_{n'}(Z')}\sqrt {R_n(Z)}}{(2\pi)^4}\!\! \int\!\! \d^3r\d^3r'
e^{-i(Y'y'-Yy+Z'z'-Zz)}\Phi_{n'}(u')\Phi_n(u)\sum_{n''}\Phi_{n''}(u') 
&  \\
& \strut\disp 
\times\!\!\!\int\!\!\d Y''\d Z''e^{-i[(y-y')Y''+(z-z')Z'']}
\rho_{\sigma n''}^{\sigma'}(\vr,Y''\!,Z'') 
\!=\!
\frac {\sqrt {R_{n'}(Z')}\sqrt {R_n(Z)}}{(2\pi)^2}\!\!\!\int\!\!\d^3r
e^{i[(Y'\!-Y)y+(Z'\!-Z)z]}\Phi_n(u)\rho_{\sigma n'}^{\sigma'}(\vr,Y'\!,Z').
& 
\nonumber
\eeq

To understand how to simplify equation (\ref{eq:res-01}), 
let us consider a 1-dimensional problem, where the electron is described by its $x$-coordinate. 
The expansion of the density matrix kernel in coordinate representation is 
given by the following expression
\be \label{eq:rhoexpanPhi}
\rho(x',x)=\sum_{n,n'}\Phi_{n'}(x')\Phi_n(x)\rho_{n,n'}.
\ee
The inverse transformation is
\be \label{eq:rhorev}
\rho_{n,n'}=\int\d x'\d x\ \Phi_{n'}(x')\Phi_n(x)\rho(x',x).
\ee
Combining (\ref{eq:rhoexpanPhi}) and (\ref{eq:rhorev}) one  gets:
\be \label{eq:rhoxxrhoxxsim}
\rho(x',x)=\sum_{n,n'}\Phi_{n'}(x')\Phi_n(x)\int\d x'''\d x''\Phi_{n'}(x''')
\Phi_n(x'')\rho(x''',x'')=\sum_{n'}\Phi_{n'}(x')\rho_{n'}(x),
\ee
where
\beq \label{eq:rhonxdef}
& \strut\disp \rho_{n'}(x)=\sum_{n}\Phi_n(x)\int\d x'''\d x''\Phi_{n'}(x''')
\Phi_n(x'')\rho(x''',x'')=\int\d x'''\d x''\Phi_{n'}(x''')\delta(x''-x)\rho(x''',x'')
= & \nonumber \\
& \strut\disp =\int\d x'\Phi_{n'}(x')
\rho(x',x). &
\eeq
If one considers the density matrix on a sufficiently short time scales,  
then there are no mixed states in the Landau levels and the coefficients $\rho_{n,n'}$ in equation 
(\ref{eq:rhoexpanPhi}) have only diagonal terms $\rho_{n,n'}=\delta_{n}^{n'}\rho_{n,n}$. 
Therefore, 
\be \label{eq:rhoxxrhodel}
\rho(x',x)=\sum_{n,n'}\Phi_{n'}(x')\Phi_n(x)\rho_{n,n}.
\ee
Substituting this to equation (\ref{eq:rhonxdef}), we get
\be \label{eq:rhoxxrhodel}
\rho_{n}(x)=\Phi_{n}(x)\rho_{n,n}.
\ee
This proves that in a more general case $\rho_{\sigma n'}^{\sigma'}(\vr,Y'\!,Z')=\Phi_{n'}(u)\rho_{\sigma n}^{\sigma'}(Z')$.
This simplifies relation (\ref{eq:res-01}):
\beq \label{eq:result}
& \strut\disp \rho_{\sigma n}^{\sigma' n'}\left({Y'Z'}\atop {YZ}\right)=
\frac {\sqrt {R_{n'}(Z')}\sqrt {R_n(Z)}}{(2\pi)^2}\int\d^3re^{i[(Y'-Y)y+(Z'-Z)z]}
\Phi_n(u)\Phi_{n'}(u)\rho_{\sigma n'}^{\sigma'}(Z)= & \nonumber \\
& \strut\disp =\sqrt {R_{n'}(Z')}\sqrt {R_n(Z)}\delta(Y'-Y)\delta(Z'-Z)\delta_{n}^{n'}
\rho_{\sigma n'}^{\sigma'}(Z')=R_n(Z)\delta(Y'-Y)\delta(Z'-Z)\delta_{n}^{n'}
\rho_{\sigma n}^{\sigma'}(Y,Z). &
\eeq
This completes the proof of equation (\ref{eq:rhoRnZ}).

\section{Expressions for matrices in equations (\ref{eq:Sp_p_I1})--(\ref{eq:Sp_p_I3})}
\label{app_st1}

The expressions for the matrix  $\hat{\cF}=\{F_{ij}\}$ in equation (\ref{eq:Sp_p_I1}) can be written in the following form: 
\beq \label{eq:matr_f}
& \strut\disp 
F_{1k}=F_{k1}=F_{mm}=0, \quad (k,m=1,2,3,4), &  \\  
& F_{23}=\langle^1_2\rangle-\langle^2_1\rangle,\quad F_{24}=-i(\langle^1_2\rangle
+\langle^2_1\rangle),\quad F_{32}=-F_{23},\quad F_{34}=i(\langle^1_1\rangle-
\langle^2_2\rangle),\quad F_{42}=-F_{24},\quad F_{43}=-F_{34}, \nonumber
& 
\eeq
where we defined 
\beq \label{eq:mark1}
\langle^{j}_{k}\rangle\equiv M_{\sigma k}^{\sigma'j}\left({{nYZ}\atop {nYZ}}\right.
\left|{{\vk_1}\atop {\vk_1}}\right) .
\eeq
The elements of matrices $\hat{\cF}'=\{F'_{ij}\}$ and $\hat{\cF}'=\{F''_{ij}\}$ can be represented as 
\beq \label{eq:f'}
& \strut\disp \left(^{F'_{11}}_{F'_{21}}\right)=2(\gamma_{11}^{11}+\gamma_{21}^{12}\pm\gamma_{12}^{21}\pm\gamma_{22}^{22}),\quad
\left(^{F'_{12}}_{F'_{22}}\right)=2(\gamma_{11}^{11}-\gamma_{21}^{12}\pm\gamma_{12}^{21}\mp\gamma_{22}^{22}), & \nonumber \\
& \strut\disp \left(^{F'_{13}}_{F'_{23}}\right)=2(\gamma_{11}^{12}+\gamma_{21}^{11}\pm\gamma_{12}^{22}\pm\gamma_{22}^{21}),\quad
\left(^{F'_{14}}_{F'_{24}}\right)=2i(\gamma_{11}^{12}-\gamma_{21}^{11}\pm\gamma_{12}^{22}\mp\gamma_{22}^{21}), & \nonumber \\
& \strut\disp \left(^{F'_{31}}_{F'_{41}}\right)=2\left(^1_i\right)(\gamma_{11}^{21}+\gamma_{21}^{22}\pm\gamma_{12}^{11}\pm\gamma_{22}^{12}),\quad
\left(^{F'_{32}}_{F'_{42}}\right)=2\left(^{1}_{i}\right)(\gamma_{11}^{21}-\gamma_{21}^{22}\pm\gamma_{12}^{11}\mp\gamma_{22}^{12}),& \\
& \strut\disp \left(^{F'_{33}}_{F'_{43}}\right)=2\left(^{1}_{i}\right)(\gamma_{11}^{22}+\gamma_{21}^{21}\pm\gamma_{12}^{12}\pm\gamma_{22}^{11}),\quad
\left(^{F'_{34}}_{F'_{44}}\right)=2\left({_{-}}{^{i}_{1}}\right)(\gamma_{11}^{22}-\gamma_{21}^{21}\pm\gamma_{12}^{12}\mp\gamma_{22}^{11}), & \nonumber 
\eeq
\beq\label{eq:f''}
& \strut\disp 
F''_{jj}=-\gamma_{s1}^{1s}-\zeta_{s1}^{1s}-\gamma_{s2}^{2s}-\zeta_{s2}^{2s} & \quad (j=1,2,3,4),
 \nonumber \\
& \strut\disp 
F''_{12}=F''_{21}=-\gamma_{s1}^{1s}-\zeta_{s1}^{1s}+\gamma_{s2}^{2s}+\zeta_{s2}^{2s},  & \quad 
F''_{13}=F''_{31}=-\gamma_{s2}^{1s}-\zeta_{s1}^{2s}-\gamma_{s1}^{2s}-\zeta_{s2}^{1s},
 \nonumber \\
& \strut\disp 
F''_{14}=F''_{41}=i(\gamma_{s2}^{1s}-\zeta_{s1}^{2s}-\gamma_{s1}^{2s}+\zeta_{s2}^{1s}), & \quad
F''_{23}=-F''_{32}=-\gamma_{s2}^{1s}-\zeta_{s1}^{2s}+\gamma_{s1}^{2s}+\zeta_{s2}^{1s},  \\
& \strut\disp 
F''_{24}=-F''_{42}=i(\gamma_{s2}^{1s}-\zeta_{s1}^{2s}+\gamma_{s1}^{2s}-\zeta_{s2}^{1s}), &\quad 
F''_{34}=-F''_{43}=i(-\gamma_{s1}^{1s}+\zeta_{s1}^{1s}+\gamma_{s2}^{2s}-\zeta_{s2}^{2s}) ,  \nonumber 
\eeq
where
\beq \label{eq:mark2}
\strut\disp
\gamma_{km}^{jl}\equiv
M_{\sigma k}^{\sigma' j}\left({{nYZ}\atop {nYZ}}\left|{{\vk_1}\atop {\vk}}\right)
M_{\sigma'' m}^{\sigma''' l}\left({{n'Y'Z'}\atop {n'Y'Z'}}\right|{{\vk}\atop {\vk_1}}\right),\quad
\zeta_{km}^{jl}\equiv 
M_{\sigma'' k}^{\sigma''' j}\left({{n'Y'Z'}\atop {n'Y'Z'}}\left|{{\vk_1}\atop {\vk}}\right)
M_{\sigma m}^{\sigma' l}\left({{nYZ}\atop {nYZ}}\right|{{\vk}\atop {\vk_1}}\right).
\eeq
 
The matrix $\hat{\cR}$ is a sum of the products of the elements of vector $\vN$ (the Stokes parameters) and four matrices 
\be \label{eq:mart_R}
\hat{\cR}=\hat{\cR}_{\rm I} n_{\rm I}+\hat{\cR}_{\rm Q} n_{\rm Q}+\hat{\cR}_{\rm U} n_{\rm U}+\hat{\cR}_{\rm V}
n_{\rm V}.
\ee
The elements of the matrices $\hat{\cR}_{I,Q,U,V}$ are:
\beq \label{Reqpe1}
& \strut\disp \left(^{R_{11}}_{R_{12}}\right)^{(I)}=A_{11}^{11}\pm A_{12}^{21}+A_{21}^{12}\pm
A_{22}^{22},\quad
\left(^{R_{13}}_{R_{14}}\right)^{(I)}=\left(^{1}_{i}\right)(\pm A_{12}^{11}+A_{11}^{21}
\pm A_{22}^{12}+A_{21}^{22}), & \nonumber \\
& \strut\disp \left(^{R_{21}}_{R_{22}}\right)^{(I)}=A_{11}^{11}\mp A_{12}^{21}+A_{21}^{12}\mp
A_{22}^{22},\quad
\left(^{R_{23}}_{R_{24}}\right)^{(I)}=\left(^{1}_{i}\right)(B_{11}^{21}\mp B_{12}^{11}+
B_{21}^{22}\mp B_{22}^{12}), & \nonumber \\
& \strut\disp \left(^{R_{31}}_{R_{32}}\right)^{(I)}=\pm C_{11}^{21}+ C_{12}^{11}\pm
C_{21}^{22}+ C_{22}^{12},\quad
\left(^{R_{33}}_{R_{34}}\right)^{(I)}=\left(^{1}_{i}\right)(\pm C_{11}^{11}+
C_{12}^{21}\pm C_{21}^{12}+ C_{22}^{22}), & \nonumber \\
& \strut\disp \left(^{R_{41}}_{R_{42}}\right)^{(I)}=\pm D_{11}^{21}+C_{11}^{11}\pm
D_{21}^{22}+ C_{21}^{12},\quad
\left(^{R_{43}}_{R_{44}}\right)^{(I)}=\left(^{1}_{i}\right)(\mp D_{12}^{11}-
C_{12}^{21}\mp D_{22}^{12}- C_{22}^{22}), & 
\eeq
\beq
& \strut\disp \left(^{R_{11}}_{R_{12}}\right)^{(Q)}=A_{11}^{11}\pm A_{12}^{21}-A_{21}^{12}\mp
A_{22}^{22},\quad
\left(^{R_{13}}_{R_{14}}\right)^{(Q)}=\left(^{1}_{i}\right)(\pm A_{12}^{11}+
A_{11}^{21}\mp A_{22}^{12}-A_{21}^{22}), & \nonumber \\
& \strut\disp \left(^{R_{21}}_{R_{22}}\right)^{(Q)}=A_{11}^{11}\mp A_{12}^{21}-A_{21}^{12}\pm
A_{22}^{22},\quad
\left(^{R_{23}}_{R_{24}}\right)^{(Q)}=\left(^{1}_{i}\right)(B_{11}^{21}\mp B_{12}^{11}-
B_{21}^{22}\pm B_{22}^{12}), & \nonumber \\
& \strut\disp \left(^{R_{31}}_{R_{32}}\right)^{(Q)}=\pm C_{11}^{21}+ C_{12}^{11}\mp
C_{21}^{22}-C_{22}^{12},\quad
\left(^{R_{33}}_{R_{34}}\right)^{(Q)}=\left(^{1}_{i}\right)(\pm C_{11}^{11}+
C_{12}^{21}\mp C_{21}^{12}- C_{22}^{22}), & \nonumber \\
& \strut\disp \left(^{R_{41}}_{R_{42}}\right)^{(Q)}=\pm D_{11}^{21}+ C_{11}^{11}\mp
D_{22}^{22}-C_{21}^{12},\quad
\left(^{R_{43}}_{R_{44}}\right)^{(Q)}=\left(^{1}_{i}\right)(\mp D_{12}^{11}-
C_{12}^{21}\pm D_{22}^{12}+C_{22}^{22}), & 
\eeq
\beq
& \strut\disp \left(^{R_{11}}_{R_{12}}\right)^{(U)}=A_{21}^{11}\pm A_{22}^{21}+A_{11}^{12}\pm
A_{12}^{22},\quad
\left(^{R_{13}}_{R_{14}}\right)^{(U)}=\left(^{1}_{i}\right)(\pm A_{22}^{11}+
A_{21}^{21}\pm A_{12}^{12}+A_{11}^{22}), & \nonumber \\
& \strut\disp \left(^{R_{21}}_{R_{22}}\right)^{(U)}=A_{21}^{11}\mp A_{22}^{21}+A_{11}^{12}\mp
A_{12}^{22},\quad
\left(^{R_{23}}_{R_{24}}\right)^{(U)}=\left(^{1}_{i}\right)(B_{21}^{21}\mp B_{22}^{11}+
B_{11}^{22}\mp B_{12}^{12}), & \nonumber \\
& \strut\disp \left(^{R_{31}}_{R_{32}}\right)^{(U)}=\pm C_{21}^{21}+ C_{22}^{11}\pm
C_{11}^{22}+C_{12}^{12},\quad
\left(^{R_{33}}_{R_{34}}\right)^{(U)}=\left(^{1}_{i}\right)(\pm C_{21}^{11}+
C_{22}^{21}\pm C_{11}^{12}+C_{12}^{22}), & \nonumber \\
& \strut\disp \left(^{R_{41}}_{R_{42}}\right)^{(U)}=\pm D_{21}^{21}+C_{21}^{11}\pm
D_{11}^{22}+C_{11}^{12},\quad
\left(^{R_{43}}_{R_{44}}\right)^{(U)}=\left(^{1}_{i}\right)(\mp D_{22}^{11}-
C_{22}^{21}\mp D_{12}^{12}-C_{12}^{22}), & 
\eeq
\beq\label{Reqpe2}
& \strut\disp \left(^{R_{11}}_{R_{12}}\right)^{(V)}=i(-A_{21}^{11}\mp A_{22}^{21}+A_{11}^{12}
\pm A_{12}^{22}),\quad
\left(^{R_{13}}_{R_{14}}\right)^{(V)}=\left(^{i}_{1}\right)(-A_{22}^{11}\mp A_{21}^{21}+
A_{12}^{12}\pm A_{11}^{22}), & \nonumber \\
& \strut\disp \left(^{R_{21}}_{R_{22}}\right)^{(V)}=i(-A_{21}^{11}\pm A_{22}^{21}+A_{11}^{12}
\mp A_{12}^{22}),\quad
\left(^{R_{23}}_{R_{24}}\right)^{(V)}=\left(^{i}_{1}\right)(\mp B_{21}^{21}+B_{22}^{11}
\pm B_{11}^{22}- B_{12}^{12}), & \nonumber \\
& \strut\disp \left(^{R_{31}}_{R_{32}}\right)^{(V)}=i(\mp C_{21}^{21}- C_{22}^{11}\pm
C_{11}^{22}+ C_{12}^{12}),\quad
\left(^{R_{33}}_{R_{34}}\right)^{(V)}=\left(^{i}_{1}\right)(-C_{21}^{11}\mp C_{22}^{21}+
C_{11}^{12}\pm C_{12}^{22}), & \nonumber \\
& \strut\disp \left(^{R_{41}}_{R_{42}}\right)^{(V)}=i(\mp D_{21}^{21}- C_{21}^{11}\pm
D_{11}^{22}+ C_{11}^{12}),\quad
\left(^{R_{43}}_{R_{44}}\right)^{(V)}=\left(^{i}_{1}\right)(D_{22}^{11}\pm C_{22}^{21}-
D_{12}^{12}\mp C_{12}^{22}) , &
\eeq
where we introduced the following combinations of the scattering matrices: 
\be
A_{km}^{jl}=
M_{\sigma'' k}^{\sigma''' j}\left({{n'Y'Z'}\atop {nYZ}}\left|{{\vk_1}\atop {\vk}}\right)
M_{\sigma m}^{\sigma'l}\left({{nYZ}\atop {n'Y'Z'}}\right|{{\vk}\atop {\vk_1}}\right)
+
M_{\sigma'k}^{\sigma j}\left({{n'Y'Z'}\atop {nYZ}}\left|{{\vk_1}\atop {\vk}}\right)
M_{\sigma'''m}^{\sigma''l}\left({{nYZ}\atop{n'Y'Z'}}\right|{{\vk}\atop {\vk_1}}\right),
\ee
\be 
B_{km}^{jl}=
M_{\sigma'' k}^{\sigma''' j}\left({{n'Y'Z'}\atop {nYZ}}\left|{{\vk_1}\atop {\vk}}\right)
M_{\sigma m}^{\sigma'l}\left({{nYZ}\atop {n'Y'Z'}}\right|{{\vk}\atop {\vk_1}}\right)
-
M_{\sigma'k}^{\sigma j}\left({{n'Y'Z'}\atop {nYZ}}\left|{{\vk_1}\atop {\vk}}\right)
M_{\sigma'''m}^{\sigma''l}\left({{nYZ}\atop{n'Y'Z'}}\right|{{\vk}\atop {\vk_1}}\right),
\ee
\be 
C_{km}^{jl}=
M_{\sigma'' k}^{\sigma''' j}\left({{n'Y'Z'}\atop {nYZ}}\left|{{\vk_1}\atop {\vk}}\right)
M_{\sigma m}^{\sigma'l}\left({{nYZ}\atop {n'Y'Z'}}\right|{{\vk}\atop {\vk_1}}\right)
+
M_{\sigma'k}^{\sigma (3-j)}\left({{n'Y'Z'}\atop {nYZ}}\left|{{\vk_1}\atop {\vk}}\right)
M_{\sigma'''(3-m)}^{\sigma''l}\left({{nYZ}\atop{n'Y'Z'}}\right|{{\vk}\atop {\vk_1}}\right),
\ee
\be 
D_{km}^{jl}=
M_{\sigma'' k}^{\sigma''' j}\left({{n'Y'Z'}\atop {nYZ}}\left|{{\vk_1}\atop {\vk}}\right)
M_{\sigma m}^{\sigma'l}\left({{nYZ}\atop {n'Y'Z'}}\right|{{\vk}\atop {\vk_1}}\right)
-
M_{\sigma'k}^{\sigma (3-j)}\left({{n'Y'Z'}\atop {nYZ}}\left|{{\vk_1}\atop {\vk}}\right)
M_{\sigma'''(3-m)}^{\sigma''l}\left({{nYZ}\atop{n'Y'Z'}}\right|{{\vk}\atop {\vk_1}}\right) . 
\ee

The elements of the matrices $\hat{\cR}'$ are 
\beq
& \strut\disp 
\left(^{R'_{11}}_{R'_{12}}\right)=\xi_{11}^{11}+\xi_{21}^{12}\pm\xi_{12}^{21}\pm\xi_{22}^{22},\quad 
\left(^{R'_{13}}_{R'_{14}}\right)=\left(^{1}_{i}\right)\left(\xi_{12}^{11}+\xi_{22}^{12}\pm\xi_{11}^{21}\pm\xi_{21}^{22}\right), & \nonumber \\
& \strut\disp 
\left(^{R'_{21}}_{R'_{22}}\right)=\xi_{11}^{11}-\xi_{12}^{21}\pm\xi_{21}^{12}\mp\xi_{22}^{22},\quad \left(^{R'_{23}}_{R'_{24}}\right)=\left(^{1}_{i}\right)
\left(\xi_{21}^{11}-\xi_{22}^{21}+\xi_{11}^{12}-\xi_{12}^{22}\right), & \\
& \strut\disp 
\left(^{R'_{31}}_{R'_{32}}\right)=\xi_{11}^{21}+\xi_{12}^{11}\pm\xi_{21}^{22}\pm\xi_{22}^{12},\quad 
\left(^{R'_{33}}_{R'_{34}}\right)=\left(^{1}_{i}\right)\left(\pm\xi_{21}^{21}\pm\xi_{22}^{11}+\xi_{11}^{22}+\xi_{12}^{12}\right), & \nonumber \\
& \strut\disp 
\left(^{R'_{41}}_{R'_{42}}\right)=\xi_{11}^{21}-\xi_{12}^{11}\pm\xi_{21}^{22}\mp\xi_{22}^{12},\quad
\left(^{R'_{43}}_{R'_{44}}\right)=\left(^{1}_{i}\right)\left(\pm\xi_{21}^{21}\mp\xi_{22}^{11}+\xi_{11}^{22}-\xi_{12}^{12}\right), & \nonumber
\eeq
where we defined 
\beq
\xi_{km}^{jl}\equiv
& \strut\disp
M_{\sigma k}^{\sigma' j}\left({{n'Y'Z'}\atop {nYZ}}\left|{{\vk_1}\atop {\vk}}\right)
M_{\sigma'' m}^{\sigma''' l}\left({{nYZ}\atop {n'Y'Z'}}\right|{{\vk}\atop {\vk_{1}}}\right) .
\eeq
The elements of the matrices $\hat{\cR}''$ are 
\beq
& \strut\disp 
\left(^{R''_{11}}_{R''_{12}}\right)=\phi_{s1}^{1s}+\varrho_{s1}^{1s}\pm\phi_{s2}^{2s}\pm\varrho_{s2}^{2s},\quad
\left(^{R''_{13}}_{R''_{14}}\right)=\left(^{1}_{i}\right)\left(\pm\phi_{s2}^{1s}\pm\varrho_{s2}^{1s}+\phi_{s1}^{2s}+\varrho_{s1}^{2s}\right), 
& \nonumber \\
& \strut\disp 
\left(^{R''_{21}}_{R''_{22}}\right)=\left(^{R''_{12}}_{R''_{11}}\right),\quad
\left(^{R''_{23}}_{R''_{24}}\right)=\left(^{1}_{i}\right)\left(\mp\phi_{s2}^{1s}\pm\varrho_{s2}^{1s}+\phi_{s1}^{2s}-\varrho_{s1}^{2s}\right), 
& \\
& \strut\disp 
\left(^{R''_{31}}_{R''_{32}}\right)=\phi_{s2}^{1s}\pm\varrho_{s2}^{1s}\pm\phi_{s1}^{2s}+\varrho_{s1}^{2s},\quad
\left(^{R''_{33}}_{R''_{34}}\right)=\left(^{1}_{i}\right)\left(\pm\phi_{s1}^{1s}+\varrho_{s1}^{1s}+\phi_{s2}^{2s}\pm\varrho_{s2}^{2s}\right), 
& \nonumber \\
& \strut\disp 
\left(^{R''_{41}}_{R''_{42}}\right)=\left({^{}_{-}}{^{R''_{14}}_{R''_{24}}}\right),\quad
\left(^{R''_{43}}_{R''_{44}}\right)=\left({^{-}}{^{R''_{34}}_{R''_{11}}}\right) , \nonumber &
\eeq
where we defined 
\beq
\phi_{km}^{jl}\equiv
M_{\sigma'' k}^{\sigma''' j}\left({{n'Y'Z'}\atop {nYZ}}\left|{{\vk_1}\atop {\vk}}\right)
M_{\sigma m}^{\sigma' l}\left({{nYZ}\atop {n'Y'Z'}}\right|{{\vk}\atop {\vk_{1}}}\right),\quad  
\varrho_{km}^{jl}\equiv
M_{\sigma' k}^{\sigma''' j}\left({{n'Y'Z'}\atop {nYZ}}\left|{{\vk_1}\atop {\vk}}\right)
M_{\sigma m}^{\sigma'' l}\left({{nYZ}\atop {n'Y'Z'}}\right|{{\vk}\atop {\vk_{1}}}\right). 
\eeq

\section{Expressions for matrices in equations (\ref{eq:Sp_np_J1})--(\ref{eq:Sp_np_J3})}
\label{app_st2}

Matrix $\hat{\cF}=\{F_{ij}\}$ in equation  (\ref{eq:Sp_np_J1})  has the same form as in the case of polarized electrons (\ref{eq:matr_f}), but where $\sigma'=\sigma$ should be substituted in the definitions of $\langle^j_k\rangle$. 
Matrices $\hat{\cF}'$ and $\hat{\cF}''$ also have the same forms as in the case of polarized electrons given by equations (\ref{eq:f'}) and (\ref{eq:f''}),  where $\sigma'=\sigma$ and $\sigma'''=\sigma''$  should be used in the expression for $\gamma _{jm}^{ik}$ and $\zeta _{jm}^{ik}$.

Matrix $\hat{\cR}$ can be represented by  expression (\ref{eq:mart_R}). 
The elements of $\hat{\cR}_{I}$ can be represented through tensors $T_{jm}^{ik}$ defined by relation (\ref{T-matrix_def}):
\beq
& \strut\disp \left(^{R_{11}}_{R_{12}}\right)^{(I)}=T^{11}_{11}\pm T^{21}_{12}+T^{12}_{21}
\pm T^{22}_{22},\quad
\left(^{R_{13}}_{R_{14}}\right)^{(I)}=(^{1}_{i})(T^{21}_{11}\pm T^{11}_{12}+T^{22}_{21}\pm
T^{12}_{22}), & \nonumber \\
& \strut\disp 
\left(^{R_{21}}_{R_{22}}\right)^{(I)}=\left(^{R_{12}}_{R_{11}}\right)^{(I)},\quad
\left(^{R_{23}}_{R_{24}}\right)^{(I)}=0, & \nonumber \\
& \strut\disp R_{31}^{(I)}=R_{13}^{(I)},\quad R_{32}^{(I)}=0,\quad R_{33}^{(I)}
=R_{11}^{(I)},\quad R_{34}^{(I)}=i\left(-T^{12}_{21}-T^{22}_{22}\right), & \nonumber \\
& \strut\disp R_{41}^{(I)}=i\left(T^{21}_{11}+T^{22}_{21}\right),\quad R_{42}^{(I)}=
i\left(T^{11}_{12}+T^{12}_{22}\right),\quad R_{43}^{(I)}=0,\quad R_{44}^{(I)}=T^{11}_{11}
+T^{21}_{12}. &
\eeq
The elements of $\hat{\cR}_{Q}$ can be represented as 
\beq
& \strut\disp \left(^{R_{11}}_{R_{12}}\right)^{(Q)}=T^{11}_{11}\pm T^{21}_{12}-
T^{12}_{21}\mp T^{22}_{22},\quad
\left(^{R_{13}}_{R_{14}}\right)^{(Q)}=\left(^{1}_{i}\right)\left(T^{21}_{11}\pm T^{11}_{12}-T^{22}_{21}\mp
T^{12}_{22}\right), & \nonumber \\
& \strut\disp \left(^{R_{21}}_{R_{22}}\right)^{(Q)}=\left(^{R_{12}}_{R_{11}}\right)^{(Q)},\quad
\left(^{R_{23}}_{R_{24}}\right)^{(Q)}=0, & \nonumber \\
& \strut\disp R_{31}^{(Q)}=R_{13}^{(Q)},\quad R_{32}^{(Q)}=0,\quad
R_{33}^{(Q)}=R_{11}^{(Q)},\quad R_{34}^{(Q)}=-R_{34}^{(I)}, & \nonumber \\
& \strut\disp R_{41}^{(Q)}=i\left(T^{21}_{11}-T^{22}_{21}\right),\quad R_{42}^{(Q)}=
i\left(T^{11}_{12}-T^{12}_{22}\right),\quad R_{43}^{(Q)}=0,\quad
R_{44}^{(Q)}=R_{44}^{(I)}. &
\eeq
The elements of $\hat{\cR}_{U}$ are
\beq
& \strut\disp \left(^{R_{11}}_{R_{12}}\right)^{(U)}=T^{12}_{11}\pm T^{22}_{12}+
T^{11}_{21}\pm T^{21}_{22},\quad
\left(^{R_{13}}_{R_{14}}\right)^{(U)}=\left(^{1}_{i}\right)\left(T^{22}_{11}\pm T^{12}_{12}+T^{21}_{21}\pm
T^{11}_{22}\right), & \nonumber \\
& \strut\disp \left(^{R_{21}}_{R_{22}}\right)^{(U)}=\left(^{R_{12}}_{R_{11}}\right)^{(U)},\quad
\left(^{R_{23}}_{R_{24}}\right)^{(U)}=0, & \nonumber \\
& \strut\disp R_{31}^{(U)}=R_{13}^{(U)},\quad R_{32}^{(U)}=0,\quad
R_{33}^{(U)}=R_{11}^{(U)},\quad R_{34}^{(U)}=0, & \nonumber \\
& \strut\disp R_{41}^{(U)}=i\left(T^{22}_{11}+T^{21}_{21}\right),\quad 
R_{42}^{(U)}=
i\left(T^{12}_{12}+T^{11}_{22}\right),\quad R_{43}^{(U)}=0,\quad
R_{44}^{(U)}=T^{12}_{11}+T^{22}_{12}+T^{11}_{21}+T^{21}_{22}. &
\eeq
The elements of $\hat{\cR}_{V}$ are
\beq
& \strut\disp \left(^{R_{11}}_{R_{12}}\right)^{(V)}=T^{12}_{11}\pm T^{22}_{12}-
T^{11}_{21}\mp T^{21}_{22},\quad
\left(^{R_{13}}_{R_{14}}\right)^{(V)}=\left(^{1}_{i}\right)\left(\pm T^{22}_{11}+T^{12}_{12}\mp
T^{21}_{21}-T^{11}_{22}\right), & \nonumber \\
& \strut\disp \left(^{R_{21}}_{R_{22}}\right)^{(V)}=\left(^{R_{12}}_{R_{11}}\right)^{(V)},\quad
\left(^{R_{23}}_{R_{24}}\right)^{(V)}=0, & \nonumber \\
& \strut\disp R_{31}^{(V)}=R_{13}^{(V)},\quad R_{32}^{(V)}=0,\quad
R_{33}^{(V)}=R_{11}^{(V)},\quad R_{34}^{(V)}=0, & \nonumber \\
& \strut\disp R_{41}^{(V)}=-T^{22}_{11}+T^{21}_{21},\quad R_{42}^{(V)}=
-T^{12}_{12}+T^{11}_{22},\quad R_{43}^{(V)}=0,\quad
R_{44}^{(U)}=i\left(T^{12}_{11}+T^{22}_{12}-T^{11}_{21}-T^{21}_{22}\right). &
\eeq
The expressions for the elements of $\hat{\cR}'$ are 
\beq
& \strut\disp \left(^{R'_{11}}_{R'_{12}}\right)=T^{11}_{11}+T^{21}_{12}\pm T^{12}_{21}\pm
T^{22}_{22},\quad
\left(^{R'_{13}}_{R'_{14}}\right)=\left(^{1}_{i}\right)\left(T^{12}_{11}+T^{22}_{12}\pm T^{11}_{21}\pm
T^{21}_{22}\right), & \nonumber \\
& \strut\disp \left(^{R'_{21}}_{R'_{22}}\right)=T^{11}_{11}-T^{21}_{12}\pm T^{12}_{21}\mp
T^{22}_{22},\quad
\left(^{R'_{23}}_{R'_{24}}\right)=T^{12}_{11}-T^{22}_{12}\pm T^{11}_{21}\mp T^{21}_{22},
 & \nonumber \\
& \strut\disp \left(^{R'_{31}}_{R'_{32}}\right)=T^{21}_{11}+ T^{11}_{12}\pm T^{22}_{21}
\pm T^{12}_{22},\quad
\left(^{R'_{33}}_{R'_{34}}\right)=\left(^{1}_{i}\right)\left(T^{22}_{11}+T^{12}_{12}\pm T^{21}_{21}\pm
T^{11}_{22}\right), & \nonumber \\
& \strut\disp \left(^{R'_{41}}_{R'_{42}}\right)=i\left(T^{21}_{11}-T^{11}_{12}\pm T^{22}_{21}
\mp T^{12}_{22}\right),\quad
\left(^{R'_{43}}_{R'_{44}}\right)=\left(^{i}_{1}\right)\left(\pm T^{22}_{11}\mp T^{12}_{12}+T^{21}_{21}-
T^{11}_{22}\right), &
\eeq
and the elements of $\hat{\cR}''$  are as follows: 
\beq
& \strut\disp \left(^{R''_{11}}_{R''_{12}}\right)=T^{1s}_{s1}\pm T^{2s}_{s2},\quad
\left(^{R''_{13}}_{R''_{14}}\right)=\left(^{1}_{i}\right)\left(T^{2s}_{s1}\pm T^{1s}_{s2}\right),\quad
\left(^{R''_{21}}_{R''_{22}}\right)=\left(^{R''_{12}}_{R''_{11}}\right),\quad
\left(^{R''_{23}}_{R''_{24}}\right)=0, & \nonumber \\
& \strut\disp R''_{31}=R''_{13},\quad R''_{32}=R''_{34}=R''_{42}=R''_{43}=0,
\quad R''_{33}=R''_{11}, \quad 
R''_{41}=R''_{14},\quad R''_{44}=R''_{11}. &
\eeq

\end{document}